\documentclass[aps,prd,linenumber,twocolumn,nomerge,superscriptaddress,preprintnumbers,superscriptaddress,nofootinbib,floatfix,showkeys]{revtex4-2}

\usepackage{graphicx}
\usepackage{xcolor}
\usepackage{orcidlink}
\usepackage{dcolumn}
\usepackage{hyperref}
\usepackage{amssymb}
\usepackage{xspace}
\usepackage{amsmath}
\usepackage{soul}
\usepackage[mathlines]{lineno}
\usepackage{setspace} 
\usepackage[mathlines]{lineno}

\newcommand{\hu}{\xspace \ensuremath{{\rm km \, s^{-1} \, Mpc^{-1}}\xspace}}

\newcommand{\Msol}{\xspace\rm{M}_{\odot}\xspace}

\newcommand{\snr}{\textsc{SNR}\xspace} 
\newcommand{\de}{{\rm d}}

\renewcommand{\vec}[1]{\ensuremath{\overrightarrow{#1}}}

\begin{document}

\preprint{Short-author PP}

\title{A Study of Systematics on the Cosmological Inference of the Hubble Constant from Gravitational Wave Standard Sirens}

\author{Grégoire Pierra \orcidlink{0000-0003-3970-7970}}
\email{g.pierra@ip2i.in2p3.fr}
\affiliation{Université Lyon, Université Claude Bernard Lyon 1, CNRS, IP2I Lyon/IN2P3,UMR 5822, F-69622 Villeurbanne, France}

\author{Simone Mastrogiovanni \orcidlink{0000-0003-1606-4183}}
\affiliation{INFN, Sezione di Roma, I-00185 Roma, Italy}

\author{Stéphane Perriès \orcidlink{0000-0003-2213-3579}}
\affiliation{Université Lyon, Université Claude Bernard Lyon 1, CNRS, IP2I Lyon/IN2P3,UMR 5822, F-69622 Villeurbanne, France}

\author{Michela Mapelli \orcidlink{0000-0001-8799-2548}}

\affiliation{Institut für Theoretische Astrophysik, ZAH, Universität Heidelberg, 69120 Heidelberg, Germany}

\date{\today}

\begin{abstract}

Gravitational waves (GWs) from compact binary coalescences (CBCs) can constrain the cosmic expansion of the universe. In the absence of an associated electromagnetic counterpart, the spectral sirens method exploits the relation between the detector frame and the source frame masses to jointly infer the  parameters of the mass distribution of black holes (BH) and the cosmic expansion parameter $H_0$. This technique relies on the choice of the parametrization for the source mass population of BHs observed in binary black holes merger (BBHs). Using astrophysically motivated BBH populations, we study the possible systematic effects affecting the  inferred value for $H_0$ when using heuristic mass models like a broken power law, a power law plus peak and a multi-peak distributions. 
 We find that with 2000 detected GW mergers, the resulting $H_0$ obtained with a spectral sirens analysis can be biased up to $3\sigma$. The main sources of this bias come from the failure of the heuristic mass models used so far to account for a possible redshift evolution of the mass distribution and from their inability to model unexpected mass features. We conclude that future dark siren GW cosmology analyses should make use of source mass models able to account for redshift evolution and capable to adjust to unforeseen mass features. 
\end{abstract}

\keywords{Hubble constant - Dark sirens - Cosmology - Gravitational waves - Binary black hole - Redshift evolution - Mass spectrum}
                              
\maketitle

\newpage

\section{\label{Sec1}Introduction}
September 2015 marks the date of the first detection of a gravitational-wave (GW) \cite{LIGOScientific:2016aoc} emitted by the merger of a binary black hole (BBH). Since then, the LIGO-Virgo-KAGRA scientific collaboration has detected more than 90 GW signals from compact binary coalescences (CBCs) \cite{KAGRA:2021vkt, LIGOScientific:2020ibl, LIGOScientific:2021usb}. 
These detected GW signals enable us to directly measure the luminosity distances $d_L$ of the sources without making any assumption on the cosmological model. However, GW signals do not directly provide the source redshift unless an electromagnetic (EM) counterpart is observed. The redshift is required to measure the Universe's expansion. This is an interesting prospect given the current tension on the measurements of the Universe's local expansion rate, the Hubble constant $H_0$ \cite{Freedman:2021ahq}. The multiple GW detections opened the field of GW cosmology. To exploit the tens of GW sources observed without EM counterpart, additional methods have been developed to obtain the source redshift. The ``Spectral Siren'' \cite{Ezquiaga:2022zkx} method relies on the intrinsic degeneracy between the redshift $ z$ of a source and its detector $\rm m^{\rm det}$ and source frame $\rm m^{\rm s}$ masses, through the relation $ \mathrm{m}^{\rm det} = (1+z)\mathrm{m}^{\rm s}$. The redshift of the sources is implicitly estimated by measuring the detector masses and jointly fitting the source mass distribution and cosmological parameters
\cite{Mastrogiovanni:2021wsd, Mastrogiovanni:2023emh, Mastrogiovanni:2023zbw}. This method depends on the choice of the phenomenological models for the BBH source mass distribution and their capacity to describe the true underlying BBH population \cite{Taylor:2011fs, Wysocki:2018mpo, Farr:2019twy, Mastrogiovanni:2021wsd, Mancarella:2021ecn, Mukherjee:2021rtw, Gray:2021qfw, Gray:2021sew, Gair:2022zsa, Leyde:2022orh, Ezquiaga:2022zkx}. The ``spectral sirens'' method is also tightly connected with the ``galaxy catalog method'' \cite{Schutz:1986gp,Gray:2021qfw,Gray:2021sew,Gair:2022zsa}, which identifies the possible galaxy hosts of the GW source by using galaxy surveys. Even with the extra information from galaxy surveys, BBH source mass models still have to be used. In fact, in the case that the galaxy survey is highly incomplete, the source mass models are the ones that provide the implicit redshift of the GW sources \cite{Mastrogiovanni:2023emh}.

In this study, we explore how the commonly used phenomenological source mass models can fit complex and astrophysically motivated distributions of BBH mergers. The rationale of this study is to understand how a mismatch between the source mass models and the population of BBHs would translate to an $H_0$ bias. To make this study, we start with a BBH population generated synthetically by modelling several astrophysical processes. Then we perform a hierarchical Bayesian inference \cite{Mastrogiovanni:2023zbw} to measure $H_0$ alongside the phenomenological BBH mass models. 

This paper is organized as follows. In Sec.~\ref{Sec2} we describe the framework built to simulate the GW detections, and introduce the basic principles of hierarchical Bayesian inference. We also present the phenomenological BBH population models for the mass spectrum. In Sec.~\ref{Sec3}, we discuss several sanity checks to test the validity of the framework. We also investigate the robustness of the spectral sirens method when a wrong mass model is used for the inference and when the source mass spectrum includes an unmodelled redshift evolution. In Sec.~\ref{Sec4}, we consider a spectral sirens analysis using a realistic catalog of BBH mergers, finding a biased estimate of $H_0$ if simple phenomenological mass models are employed. Sec.~\ref{Sec4} also argues about the possible different sources of the bias, specifically the importance of the redshift evolution of the mass spectrum when estimating the cosmological parameters. Finally, in Sec.~\ref{Sec5} we draw the conclusions for this study. 

\section{\label{Sec2}Bayesian inference, GW observations and CBC population modelling}

In this section, we describe our simulation procedure. We first simulate BBH observations using a detection criteria based on the signal-to-noise ratio (\snr) that we describe in Sec.~\ref{Sec2a}. Then we use the hierarchical Bayesian inference, summarized in Sec.~\ref{Sec2b}, to reconstruct cosmological and population properties.

\subsection{\label{Sec2a} A fast generator of GW observations}

The workflow that we use to generate the GW observations is the following. From a population of BBH mergers described by the two source masses and the redshift of the merger, we compute an approximate value of the optimal \snr. Instead of calculating the SNR from the full waveform and the matched filter, we use the relation below as a proxy for the optimal \snr \cite{Fishbach:2018edt, Mastrogiovanni:2021wsd}, 
\begin{equation}
 \rho = \delta \times 9 \left[\frac{\mathcal{M}_c}{\mathcal{M}_{c,{\rm 9}}} \right]^{\frac{5}{6}}\left[ \frac{d_{L,{\rm 9}}}{d_L}\right],
 \label{eq:SNR}
\end{equation}
where $\mathcal{M}_c$ is the binary detector chirp mass, $d_L$ is the luminosity distance, $\rho$ is the optimal \snr and $\delta$ a projection factor. Although Eq.~\ref{eq:SNR} can underestimate the SNR of events with chirp masses above $30 \Msol$, it should suffice to provide us with a representative sample of the BBH population.
The luminosity distance and the binary chirp mass are computed assuming a flat $\Lambda$CDM cosmology with $H_0 = 67.8$ \hu and $\Omega_{m,0}=0.308$, consistent with cosmic microwave background measurements from Planck \cite{Planck:2015fie}. The parameters with a superscript of ``$\rm 9$'' are the detector chirp mass and luminosity distance values for which a signal with $\delta=1$ would have a \snr of 9. We set the reference chirp mass to $\mathcal{M}_{c, {\rm 9}} = 25$ $\Msol$, the luminosity distance $d_{L, {\rm 9}} = 1.5$ $\rm Gpc$ \cite{KAGRA:2013rdx}. The projection factor $\delta$ captures the variations of the optimal \snr when dealing with a 3-detector network and encapsulates the combined impact of the polarization of the GWs and the sky localization. We draw it according to a cumulative density function (CDF) introduced in \cite{Dominik:2014yma}. 

To mimic the presence of detector noise in the data, we also calculate a ``\textit{detected}'' \snr $\rho^{\rm det}$ from the optimal \snr. The detected \snr is drawn from a $\chi^{2}$ distribution with $6$ degrees of freedom (2 for each detector). A GW is detected if its detected \snr exceeds a \snr threshold of $12$ and the GW frequency corresponding to the innermost stable circular orbit (ISCO) is higher than 15 Hz. The GW and ISCO frequencies are related by
\begin{equation}
 f_{\rm GW} = 2\,{}f_{\rm ISCO},
 \label{eq:frequency cut off}
\end{equation}
where $f_{\rm ISCO}$, is approximated as \cite{Moore:2016qxz}
\begin{equation}
 f_{\rm ISCO} = \frac{1}{2\pi}\frac{1}{6^{3/2}}\frac{c^3}{G} \left (\frac{1 \Msol}{M^{d}_{\rm tot}} \right )10^{3} \hspace{0.15cm}\rm Hz.
 \label{eq:f ISCO}
\end{equation}

For this study, we chose to not generate errors on the detector frame masses and luminosity distances. We assume that we are perfectly able to measure them from the GW signal. We make this choice for two motivations. The first is that we want to ease the computational load of the analysis and explore multiple test cases with thousands of GW detections. The second is that we want to maximize the potential effect of the systematics arising from the phenomenological mass models, which could be hidden by the error budget on the masses.

\begin{figure*}[ht]
    \centering
    \includegraphics[width=\textwidth]{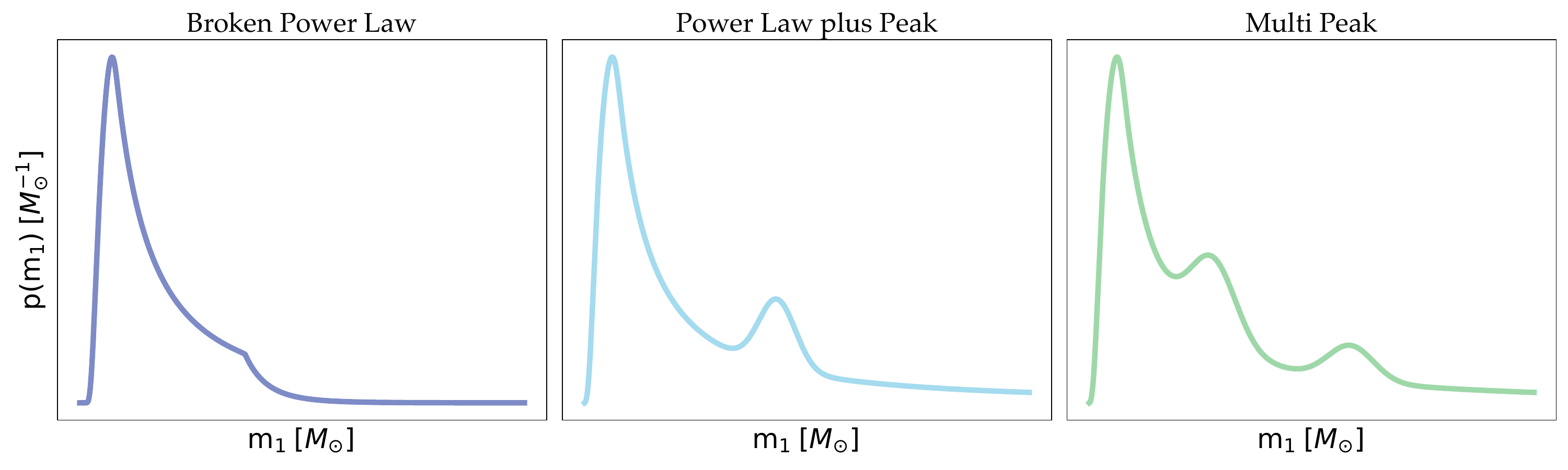}
    \caption{Simple representation of the three phenomenological mass models, the Broken Power Law in the left plot, the Power Law plus Peak in the middle plot and the Multi Peak in the right plot.}
    \label{fig:mass models checks}
\end{figure*}

\subsection{\label{Sec2b}The hierarchical likelihood}

The detection of GW events can be described by an inhomogeneous Poisson process in presence of selection biases \cite{Vitale:2020aaz, Mastrogiovanni:2023zbw, Mastrogiovanni:2023emh, Mastrogiovanni:2021wsd}. For a set of $ N_{\rm GW}$ GW signals detected over an observation time $T_{\rm obs}$, the probability of having a specific GW data set $\{\rm x \}$  given population hyper-parameters $\Lambda$ can be written as 
\begin{eqnarray}
    \mathcal{L}(\{x\}|\Lambda) &\propto &  e^{-N_{ \rm exp}(\Lambda)} \prod_i^{N_{\rm GW}} T_{\rm obs} \int \de \theta  \de z \; \times \nonumber \\ && \times \mathcal{L}_{\rm GW}(x_i|\theta,z,\Lambda) \frac{1}{1+z} \frac{\de \rm N_{ CBC}}{\de \theta \de z \de t_s}(\Lambda).
    \label{eq:likelihood}
\end{eqnarray}
The hyper-parameters $\Lambda$ describe the BBH population and also include cosmological expansion parameters. For a flat $\rm \Lambda CDM$ cosmology, the expansion parameters are the Hubble constant $H_0$ and the fraction of matter density today $\rm \Omega_{m,0}$. Other population hyper-parameters are the ones controlling the shape of the BBH mass spectrum. We describe the phenomenological models for the BBH mass spectrum in Sec.~\ref{Sec2c} and in more detail App.~\ref{App:Source mass model}. In Eq.~\ref{eq:likelihood} the individual GW likelihood $\mathcal{L}_{\rm GW}(x_i|\theta,\Lambda)$ gauges the errors on the estimation of the intrinsic parameters $\theta$ for a certain GW event $x_i$. For our study, the intrinsic parameters $\rm \theta =(m_{1}^{s},m_{2}^{s})$ denotes the masses in the source frame. In Eq.~\ref{eq:likelihood}, the factor $\frac{1}{1+ z}$ accounts for the time dilatation between the detector and source frame times $t_{\rm d}$ and $t_{\rm s}$. 
\begin{equation}
    \frac{\de \rm N_{\rm CBC}}{\de \theta \de z \de t_s} (\Lambda)
    \label{eq:cbc rate}
\end{equation}
is the CBC merger rate per source time, comoving volume and masses. Finally, $N_{\rm exp}$ is the expected number of GW detections during a time $T_{\rm obs}$. This term encapsulates the GW selection biases, and can be written as 
\begin{equation}
    N_{\rm exp} (\Lambda)= T_{\rm obs} \int \de \theta \de z \; P_{\rm det}(z,\theta,\Lambda) \frac{\de \rm N_{\rm CBC}}{\de \theta  \de z \de t_s}(\Lambda) \frac{1}{1+z},
    \label{eq:Nexp}
\end{equation}
where $P_{\rm det}(z,\theta,\Lambda)$ represents the probability of detecting a certain CBC at a redshift $z$, characterized by some specific GW parameters $\theta$ and given cosmological hyper-parameters $\Lambda$. The detection probability is given by the integral of the GW likelihood over all the detectable data realizations, namely
\begin{eqnarray}
    P_{\rm det}(z,\theta,\Lambda) = \int_{ x \in {\rm Det}} \de x \mathcal{L}_{\rm GW}(x|\theta,z,\Lambda).
    \label{eq:Pdet}
\end{eqnarray}
The Spectral Siren CBC rate is then parameterized as
\begin{eqnarray}
    \frac{\de \rm N_{\rm CBC}}{\de \theta  \de  z \de t_s} &=& \frac{\de \rm N_{\rm CBC}}{\de \vec{m}_s \de V_c \de t_s} \frac{\de V_c}{\de z }(\Lambda)\nonumber \\ &=&  R_0  \psi(z;\Lambda)  p_{\rm pop} (\vec{m}_s|\Lambda) \times \frac{\de V_c}{\de z }. \nonumber \\ && 
    \label{eq:cbc rate full}
\end{eqnarray}
In the Eq.~\ref{eq:cbc rate full}, $R_0$ is the local ($z=0$) CBC merger rate per time per comoving volume, $\psi(z;\Lambda)$ is the phenomenological function for the CBC merger rate that evolves with redshift with $\psi(z=0;\Lambda)=1$. The $\rm p_{\rm pop}$ term is a probability density function of the phenomenological models for the source frame masses $\rm p_{\rm pop}(\Vec{m}_s|\Lambda)$. The last term in Eq.~\ref{eq:cbc rate full} is the differential of the comoving volume per redshift, that for a flat $\Lambda$CDM cosmology is 
\begin{equation}
 \frac{\de V_c}{\de z} = \left[\frac{c}{H_0}\right]^3 \left[\int_0^z \frac{\de z'}{E( z')}\right]^2,
 \label{eq:diff comoving volume}
\end{equation}
where the term $E(z)$ in Eq.~\ref{eq:diff comoving volume} is the dimensionless Hubble parameter defined as 
\begin{equation}
 E(z) = \frac{H(z)}{H_0} = \sqrt{\Omega_{m}(1+z)^3+(1-\Omega_{m})}.
 \label{eq:dimensionless hubble}
\end{equation}

An important assumption that we implicitly made in Eq.~\ref{eq:cbc rate full} when writing $\rm p_{\rm pop}(\Vec{m}_s|\Lambda)$  is that the CBC mass spectrum does not evolve in redshift. This assumption is typically made in most of the state-of-the-art literature \cite{LIGOScientific:2021aug, Gray:2023wgj, Mastrogiovanni:2023zbw, Mastrogiovanni:2023emh}. In the following, we will discuss how the assumption for the source mass distribution to be independent on redshift could translate to a $H_0$ bias.

\begin{figure*}[htb!]
    \centering
    \includegraphics[scale=0.28]{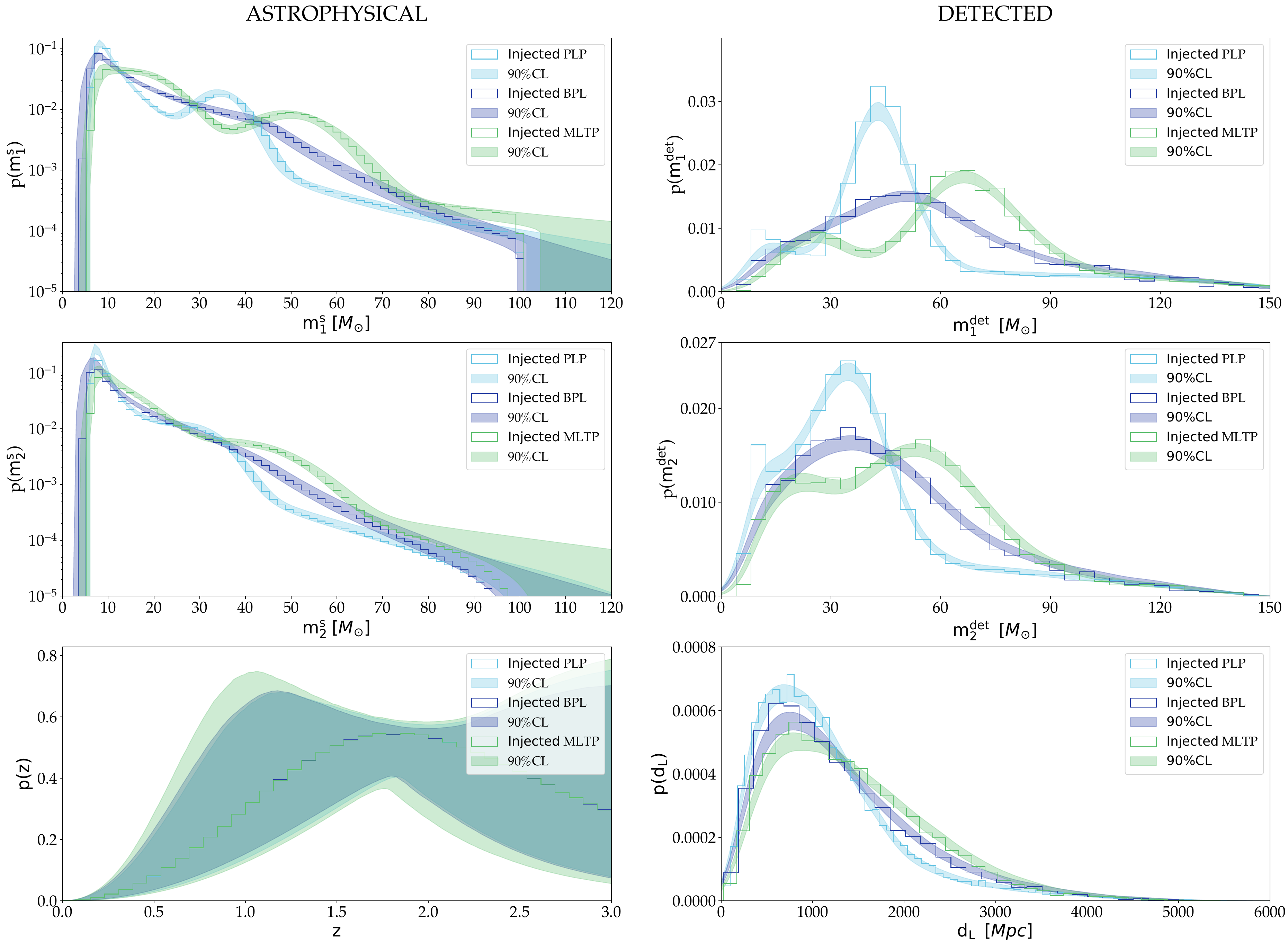}
    \caption{Posterior predictive checks for the three catalogs of BBHs simulated and inferred with the BPL (in blue), the PLP (in light blue) and the MLTP mass model (in green) with 2000 GW events. The left panels correspond to the source frame masses and redshift, from top to bottom $\rm m_1^s$, $\rm m_2^s$ and $z$. The injected populations are under the form of an histogram, and the 90\% CL reconstructed spectrum as coloured contours. The right panels displays the detector frame masses and luminosity distances, with selection effects are taken into account, from top to bottom $\rm m_1^{det}$, $\rm m_2^{det}$, $\rm d_L$.}
    \label{fig:PPC sanity checks}
\end{figure*}

\subsection{\label{Sec2c}The phenomenological population models of BBHs}

Phenomenological CBC merger rate models of masses and redshift are central to this study. Currently, spectral sirens analysis as in \cite{LIGOScientific:2021aug} use simple parametric models describing the distribution of source frame masses and CBC merger rate. For the masses, we employ three of these models: the Broken Power Law (BPL), the Power Law plus Peak (PLP) and the Multipeak (MLTP). The three mass models are represented in Fig.~\ref{fig:mass models checks}. The CBC merger rate as a function of redshift is parameterized following the Madau $\&$ Dickinson \cite{Madau:2014bja} star formation rate. In App.~\ref{App:Source mass model} and App.~\ref{App:Merger rate model}, we provide the detailed probability density functions for the models and the set of population parameters and prior ranges used in the rest of the paper. In the following, we provide a general description of these models.
\begin{figure}
    \centering
    \includegraphics[scale=0.27]{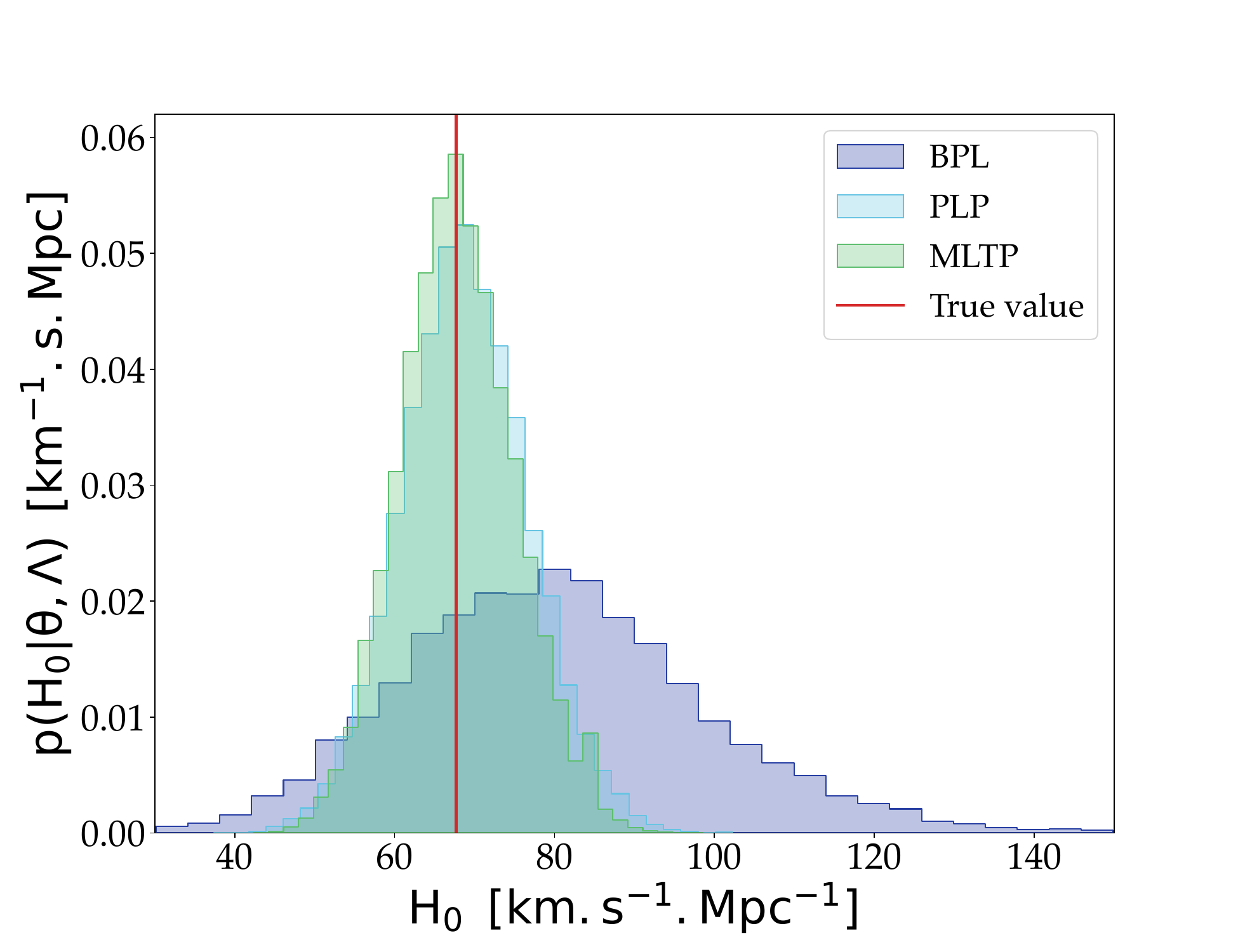}
    \caption{Marginal posterior of the Hubble constant derived from the spectral sirens inference, with the PLP, BPL and MLTP mass models and a catalog of 2000 detected GW events. The vertical red line is the injected value in the simulation, fixed to the Planck15 value of $H_0 = 67.7 \hspace{0.15cm} \hu$.}
    \label{fig:H0 sanity check}
\end{figure}

The \textit{Broken Power Law} (BPL) \cite{LIGOScientific:2020kqk} is the simplest model that we employ. It is composed of a power law, with a smooth tapering at its low mass edge plus a break in the power law at a mass $\rm m_{break}$. The low-mass smoothing is introduced to model the effects of the stellar progenitor metallicity \cite{LIGOScientific:2020kqk}, which could create a smooth transition for black hole production. The break in the power law mimics the left edge of the pair-instability supernovae \cite{Farmer:2019jed} gap. The second power law, defined after the break, could explain the existence of a second population of BBHs inside the pair-instability supernovae gap due to dynamically formed BBHs. 
\begin{figure*}[ht]
  \centering
  \includegraphics[width=\textwidth]{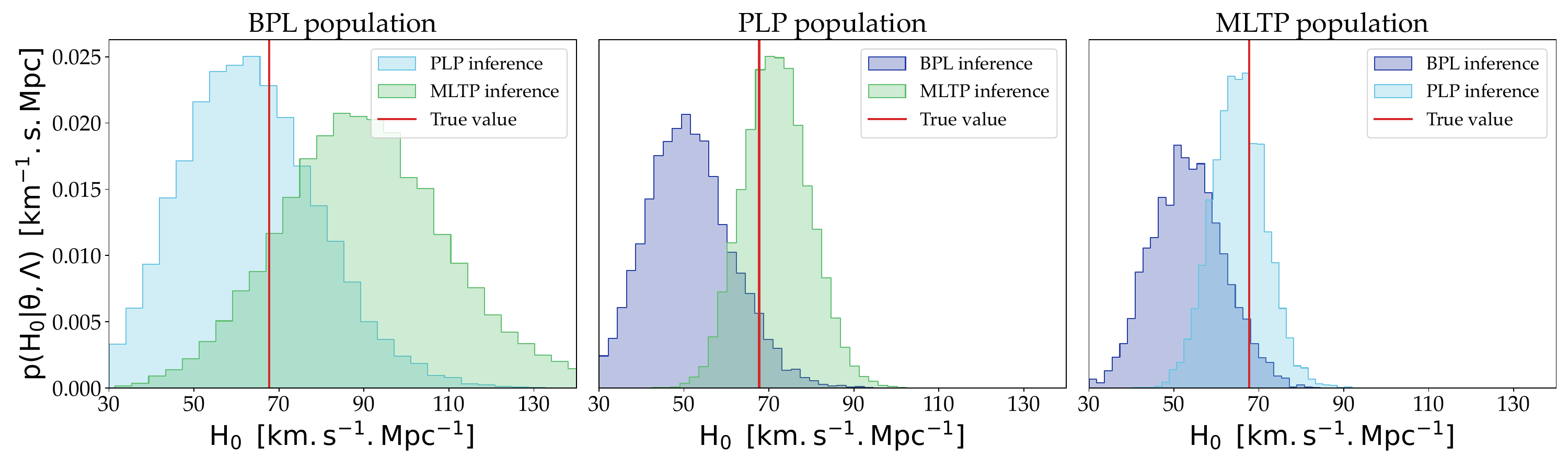}
  \caption{The left plot presents the marginal distribution of the Hubble constant inferred with the PLP and the MLTP mass models while generating the BBH catalog with the BPL model. The middle plot is produced simulating the BBH events with the PLP and inferring with the BPL and the MLTP. Finally the plot on the right shows the $H_0$ results obtained while generating the GW events with the MLTP model and inferring with the BPL and PLP mass models. On each of the plots, the true value of the Hubble constant in each population is represented by the vertical red line.}
  \label{fig:H0 wrong model checks}
\end{figure*}

The \textit{Power Law plus Peak} (PLP) \cite{Talbot:2018cva} is based on a power law with the same low mass smoothing as the BPL and incorporates a gaussian peak introduced to account for a potential accumulation of BBH just before the pair-instability supernovae gap \cite{Talbot:2018cva}. The PLP mass model is more flexible than the BPL, the position and the width of the peak can vary and allow the model to potentially catch high mass events, without changing the lower part of the spectrum. The fraction of events in the Gaussian peak is given by the parameter $\lambda_\mathrm{peak}$ (see App.~\ref{App:Source mass model}).

The \textit{Multi Peak} (MLTP) mass model is built similarly to the PLP, but with a second Gaussian peak available at higher masses. The motivations for the second Gaussian component are that BBH systems could arise from second-generation black holes produced by hierarchical mergers. The MLTP model is more flexible and can reduce into a PLP, if the secondary component is not supported by the Bayesian inference.
\begin{figure*}[htb!]
    \centering
    \includegraphics[scale=0.28]{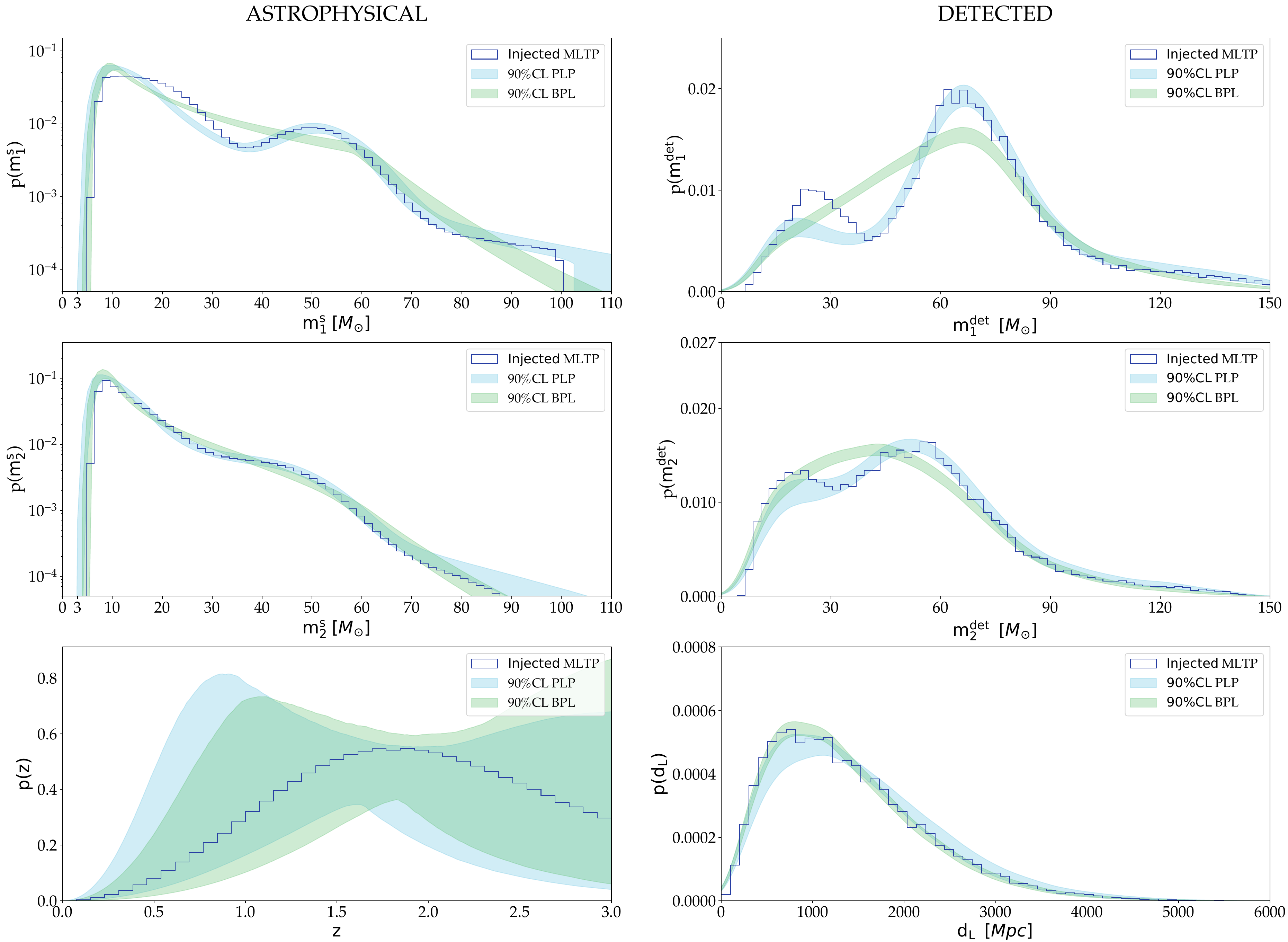}
    \caption{Posterior predictive checks for the wrong model analysis. The blue histograms are the injected population of BBHs, generated with a MLTP mass model, with the same set of parameters as depicted in Sec.~\ref{Sec3a}. The two coloured contours are the 90\% CL reconstructed populations, inferred respectively with the PLP in light blue and the BPL mass model in green. This posterior predictive check corresponds to the result presented on the right panel of Fig.~\ref{fig:H0 wrong model checks}.}
    \label{fig:PPC wrond model}
\end{figure*}

In previous studies \cite{LIGOScientific:2021aug, Mastrogiovanni:2021wsd,  Mastrogiovanni:2023emh}, a significant correlation between the Hubble constant and possible local over-densities in the BBH mass spectrum was found. In particular, the position of the gaussian peak for the PLP model ($\mu_g$) shown a strong correlation with $H_0$. 

\section{\label{Sec3} Application to analytical BBH populations}

Here we consider diverse populations of BBHs generated from parametric models to study how the PLP, BPL and MLTP models infer $H_0$ for several scenarios. In Sec.~\ref{Sec3a}, we use the same mass models (BPL, PLP and MLTP) for the generation of the BBHs catalog  for the inference. In Sec.~\ref{Sec3b}, we still generate BBHs from one of our three models, but we perform the inference using the other two. In the Sec.~\ref{Sec3c} we simulate GW events with an extended version of the PLP  that includes a redshift evolution of the Gaussian peak in order to test the bias on $H_0$ obtained by the redshift-independent models. We will quantify the $H_0$ bias according to what confidence interval (C.I.) the true values of $H_0$ are found from the hierarchical posterior. In principle, biases should be quantified using parameters-parameters (PP) plots that indicate what is the fraction of time that the true value of $H_0$ is found in a given C.I. for the posterior. As an example, if no bias is present, the true value of $H_0$ would be found in the 50\% of C.I. for 50\% of the cases and so on. However, here we do not calculate PP plots as they are computationally demanding. We argue for the presence of possible biases when the true value of $H_0$ is excluded at 2$\sigma$ or more C.I. The motivation to choose this criterion is that the probability of obtaining such discrepancy due to statistical fluctuations, and not due to biases, is only $5\%$.

In the rest of the paper, we will use the posterior predictive checks (PPCs) to assess how well the PLP, BPL and MLTP can fit the mass spectrum. The astrophysical posterior predictive distribution, given the observed GW events, for redshift and source masses is defined as
\begin{equation}
    p_{\rm pop}(z,\mathrm{m}_{1}^{\rm s},\mathrm{m}_{2}^{\rm s}|\{x\})=\int \de \Lambda \, p_{\rm pop}(z,\mathrm{m}_{1}^{\rm s},\mathrm{m}_{2}^{\rm s}|\Lambda)p(\Lambda|\{x\}).
\end{equation}
The posterior predictive distribution for the luminosity distance and detector frame masses is defined as
\begin{eqnarray}
    p_{\rm obs}(d_L,\mathrm{m}_{1}^{\de},\mathrm{m}_{2}^{\de}|\{x\})=&&\int \de \Lambda \, p_{\rm pop}(d_L,\mathrm{m}_{1}^{\de},\mathrm{m}_{2}^{\de}|\Lambda) \times \nonumber \\ &&  p(\Lambda|\{x\}) P_{\rm det}(d_L,\mathrm{m}_{1}^{\de},\mathrm{m}_{2}^{\de}).
\end{eqnarray}
We will discuss how the mapping between source and detector frame, and the mismatch of the BBH mass spectrum will translate to an $H_0$ bias.

\subsection{\label{Sec3a}Using the same mass models for simulation and inference}
We simulate three test populations of BBH mergers, with the three source mass distributions and the CBC merger rate evolution described before. The parameters of the population models used for the simulation of the catalogs are summarized in App.~\ref{App:rate pop injected}.
For the three populations, we generate 2000 detected GW events, that will be used in the spectral sirens analysis. We choose to work with 2000 GW events since for this number of detections it is typically possible to have a data-informed posterior bounded in the prior range $H_0 \in [20,140] \hu$. 
\begin{figure}
    \centering
    \includegraphics[scale=0.28]{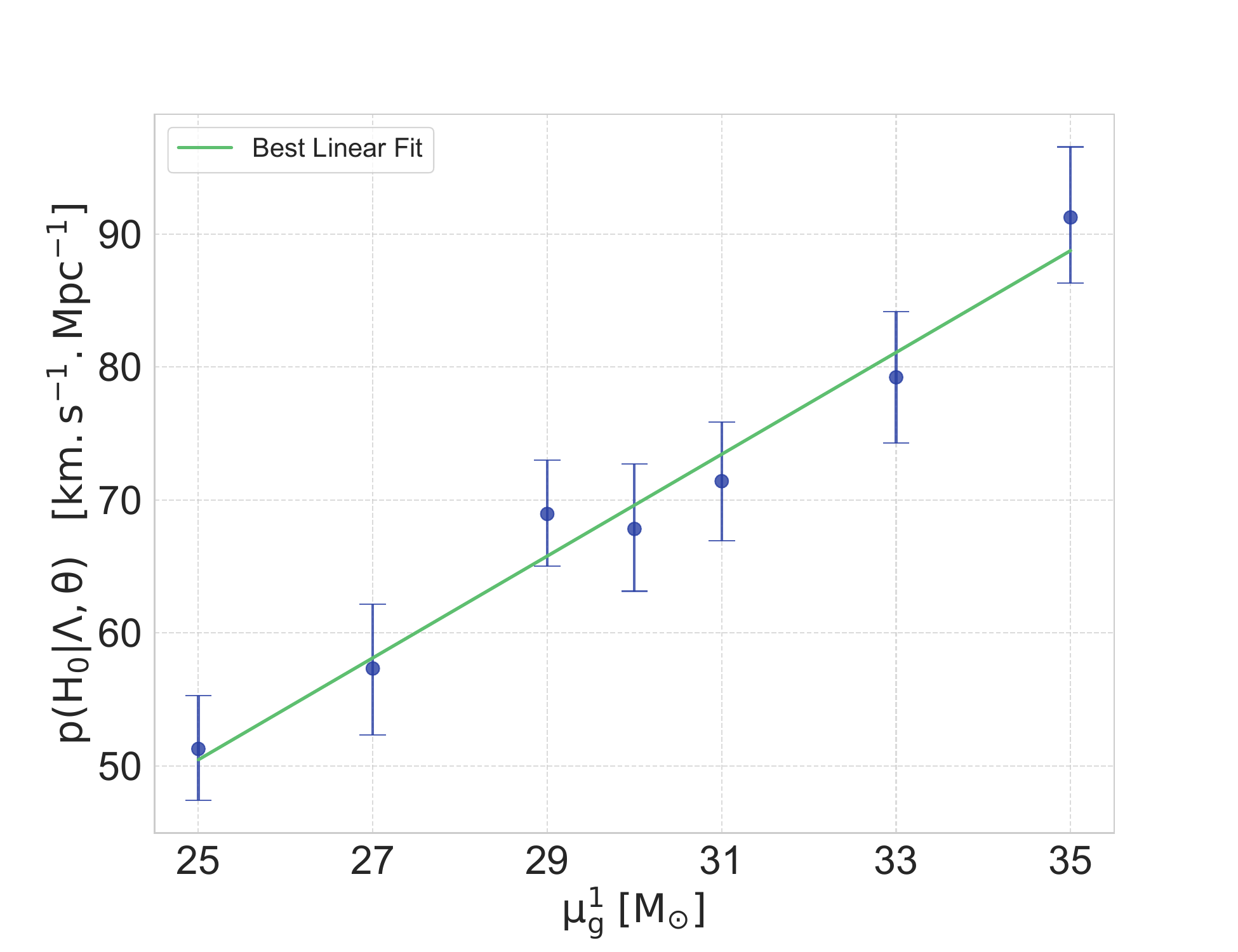}
    \caption{Evolution of the $H_0$ value inferred with respect to the final position of the PLP peak at $z=1$. The blue points correspond to the median of the inferred posterior of the Hubble constant, for each of the analysis.  The point of reference (no redshift evolution) is for $\mu_{g}=30$ at $z=1$. The error bars are the $1\sigma$ intervals of the posteriors and the green line is the best linear fit passing by all points.}
    \label{fig:H0 mug z bias}
\end{figure}
We jointly estimate the cosmological parameter $H_0$, as well as all the mass and the CBC merger rate parameters using hierarchical Bayesian inference. 

Fig.~\ref{fig:H0 sanity check} presents the marginal posterior distributions for $H_0$ obtained for each source mass model. For all the simulations, the true value of the Hubble constant is estimated within the $90\%$ C.I. We verified that the true values of all the other population parameters describing the CBC merger rate and mass spectrum are contained in the 90\% C.I. of their marginal posterior. The inferred $H_0$ is more precise for the PLP and the MLTP mass models rather than the BPL. The more precise constraint on $H_0$ is given by the fact that the PLP and MLTP contain sharper features in the source mass spectrum. 

Fig.~\ref{fig:PPC sanity checks} shows the PPCs for the BPL, PLP and MLTP mass models. The figure demonstrates that all the structures in the mass and rate spectrum are well reconstructed by the analysis. In particular, the inference can correctly reconstruct all the features present in the mass spectra. In conclusion, when using the correct mass model for inference, both the correct population and cosmological expansion parameters are recovered. Additionally, this test demonstrates the importance of having features in the BBH mass spectrum to improve the precision for cosmological expansion parameters.
\begin{figure*}[htb!]
    \centering
    \includegraphics[scale=0.28]{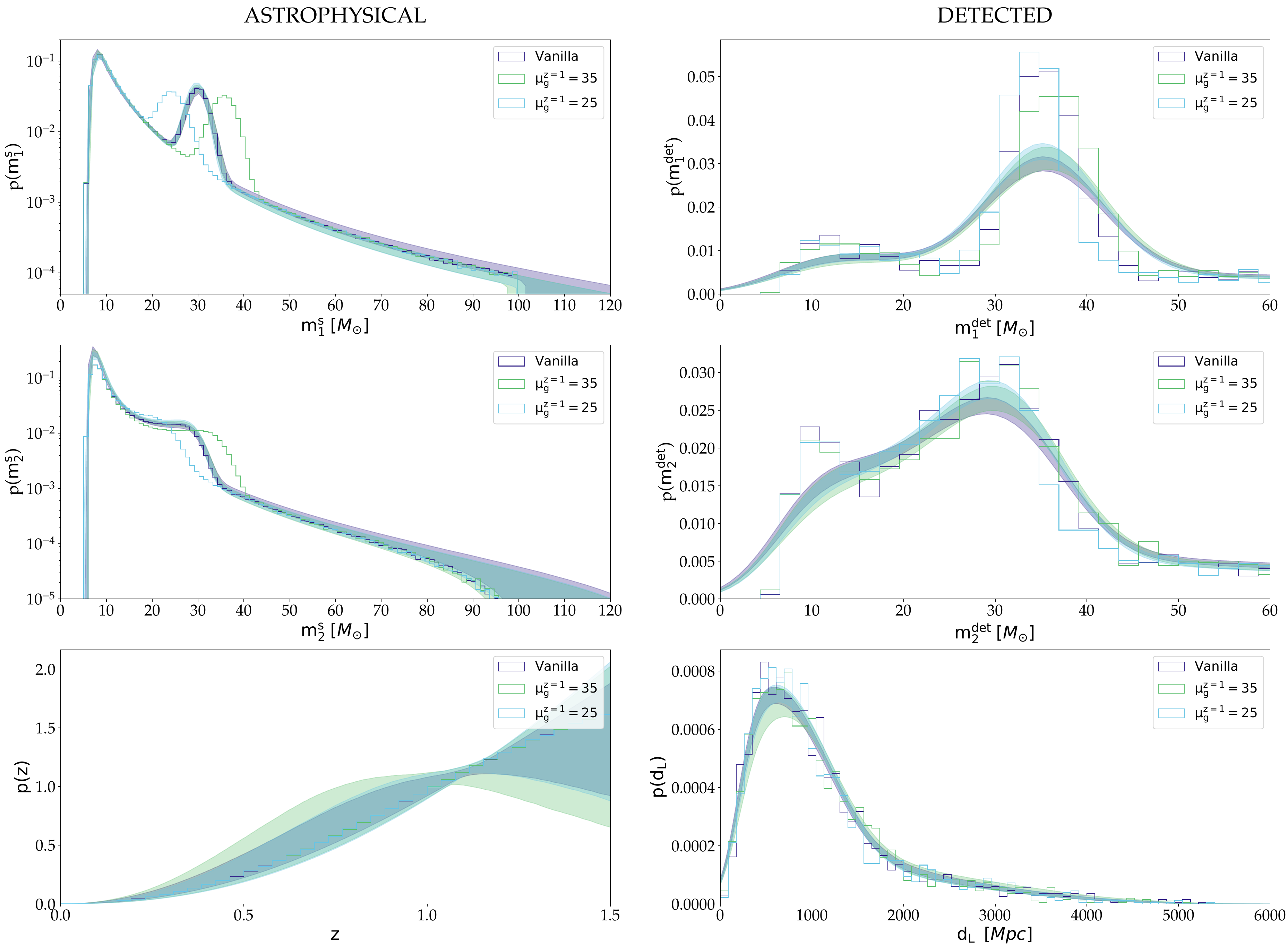}
    \caption{Posterior predictive check for the simulated populations, incorporating an additional linear evolution of the gaussian component $\mu_g(z)$ with respect of the redshift, between $z=0$ and $z=1$. The coloured contour are the 90\%CL estimated spectrum obtained with the spectral sirens analysis. The coloured histograms display the injected populations in each run, they are denoted by the position of their gaussian peak at $z=1$. Only three out of the seven populations are shown: The vanilla analysis in purple (reference run), the $\mu_{g}^{1}=35 \Msol$ in orange and the $\mu_{g}^{1}=25$ in blue.}
    \label{fig:PPC linear zevol}
\end{figure*}

\subsection{\label{Sec3b}Using different mass models for simulation and inference}

We generate 3 populations of 2000 detected  CBCs from the BPL, PLP and MLTP mass models and then we perform the estimation of populations and cosmological parameters with the other two mass models. 

Fig.~\ref{fig:H0 wrong model checks} shows the marginal posterior distributions of the Hubble constant inferred. We find that, when the detections are simulated with the BPL and the inference performed with PLP and MLTP, the true value of $\rm H_0$ is contained in the $68 \%$ C.I. despite the wrong mass model was used for the inference. In contrast, when simulating with the PLP and MLTP  and inferring with the BPL, we find the true value of $H_0$ is excluded at a 98.7\% C.I. level. This bias is linked to the difficulties of the BPL model in catching all the sharp mass features contained in the PLP and MLTP mass spectra. The mismatching of the mass features result in a systematically mismatched redshift for the sources and then on a bias on $H_0$. 

Similarly to Sec.~\ref{Sec3a}, we look at the PPC to understand the cosmological results. Fig.~\ref{fig:PPC wrond model} shows the PPCs for the case in which the simulated population was generated from a MLTP model and the Bayesian inference performed with the BPL and PLP models. From the source frame mass spectrum of $\rm m_{1}$ and $\rm m_{2}$, we see that the PLP inference can reconstruct a peak located around $50 M_{\odot}$. While the reconstructed mass spectrum from the BPL misses the high mass feature by underestimating it. The resulting bias in $H_0$ can be understood as, for lower values of $H_0$, GW events are placed at lower redshifts hence with higher source frame masses. By lowering the $H_0$ value, the BPL is trying to fit the over-density of high-mass events produced by the MLTP.
Concerning the simulations with the most complex of the models, the MLTP, we find that both the PLP and BPL are not able to correctly reconstruct the lowest part of the mass spectrum. However, the mismatch is not sufficient to introduce a $H_0$ bias for the reconstruction with the PLP. The CBC merger rate as a function of redshift, in all cases, is correctly reconstructed.  In conclusion, the sharper mass features in the BBH mass spectrum can source a significant bias on $H_0$.

\subsection{\label{Sec3c}Mass models response to a redshift evolution of the mass spectrum}

In this section, we perform a test to examine the robustness of redshift-independent mass models to a possible evolution of the mass spectrum in redshift. In doing so, we will assume a linear dependence between the features of the mass spectrum and the redshift. This simple model has already been explored in \cite{Ezquiaga:2022zkx} for the evolution of the low and high edges of the mass spectrum, finding a resulting mild $H_0$ bias when the low edge of the mass spectrum is not modelled correctly. In this paper, we modify the PLP mass model to include a peak position linearly evolving with redshift. The redshift evolution of the mass spectrum for the simulated BBH population is incorporated in the parameter that governs the position of the Gaussian component, while the edges are kept fixed in redshift. This type of evolution could be the result of an interplay of the Pair Instability Supernova mass scale and time-delays between binary formation and merger \cite{Mukherjee:2021rtw,2023A&A...677A.124K} that highly suppress the mass spectrum above the PISN scale (the gaussian peak). The impact of such models has been explored in \cite{2023MNRAS.523.4539K} on GWTC-3 finding no statistically significant preference and a consistent $H_0$ posterior against the non-evolving PLP.

Here, we do not suppress the mass spectrum above the PLP peak but just assume a linear dependency: 
\begin{equation}
    \mu_{g}(z) = \mu_{g}^{0} + z \left(\mu_{g}^{1} -\mu_{g}^{0} \right),
    \label{eq:peak evolution in z}
\end{equation}
where $\mu_{g}^{0} = \mu_{g}(z=0)$ and $\mu_{g}^{1} = \mu_{g}(z=1)$ are the positions of the gaussian peak at $z=0$ and $z=1$. We fix $\mu_{g}^{0} = 30\Msol$ and we considered 7 values for $\mu_{g}^{1}$ from $25 \Msol$ to $35 \Msol$.

For each simulation, we run a full hierarchical Bayesian inference using the redshift-independent PLP to reconstruct $H_0$ as well as the other population parameters from 2000 GW detections.
We report the $H_0$ estimation with 68.3\% C.I. for these analyses in Fig.~\ref{fig:H0 mug z bias}, showing the variations of the inferred $H_0$ value as a function of the position of $\mu_{g}^{1}$. For instance, if the Gaussian peak shifts of about 5 solar masses from redshift 0 to redshift 1, $H_0$ is found outside the 99.7\% C.I. Fig.~\ref{fig:H0 mug z bias} shows that the redshift evolution of the structures in the mass spectrum can bias the inferred value of the Hubble constant. Even when considering a mild evolution of the mass spectrum in redshift, the bias on the inferred $H_0$ is significant. We also notice that the systematic bias obtained on $H_0$ is linearly proportional to the magnitude of the redshift evolution of the mass spectrum feature. As we will describe in the following paragraph, the bias is introduced by the fact that the source-frame mass spectrum is not reconstructed correctly.
\begin{figure}
    \centering
    \includegraphics[scale=0.29]{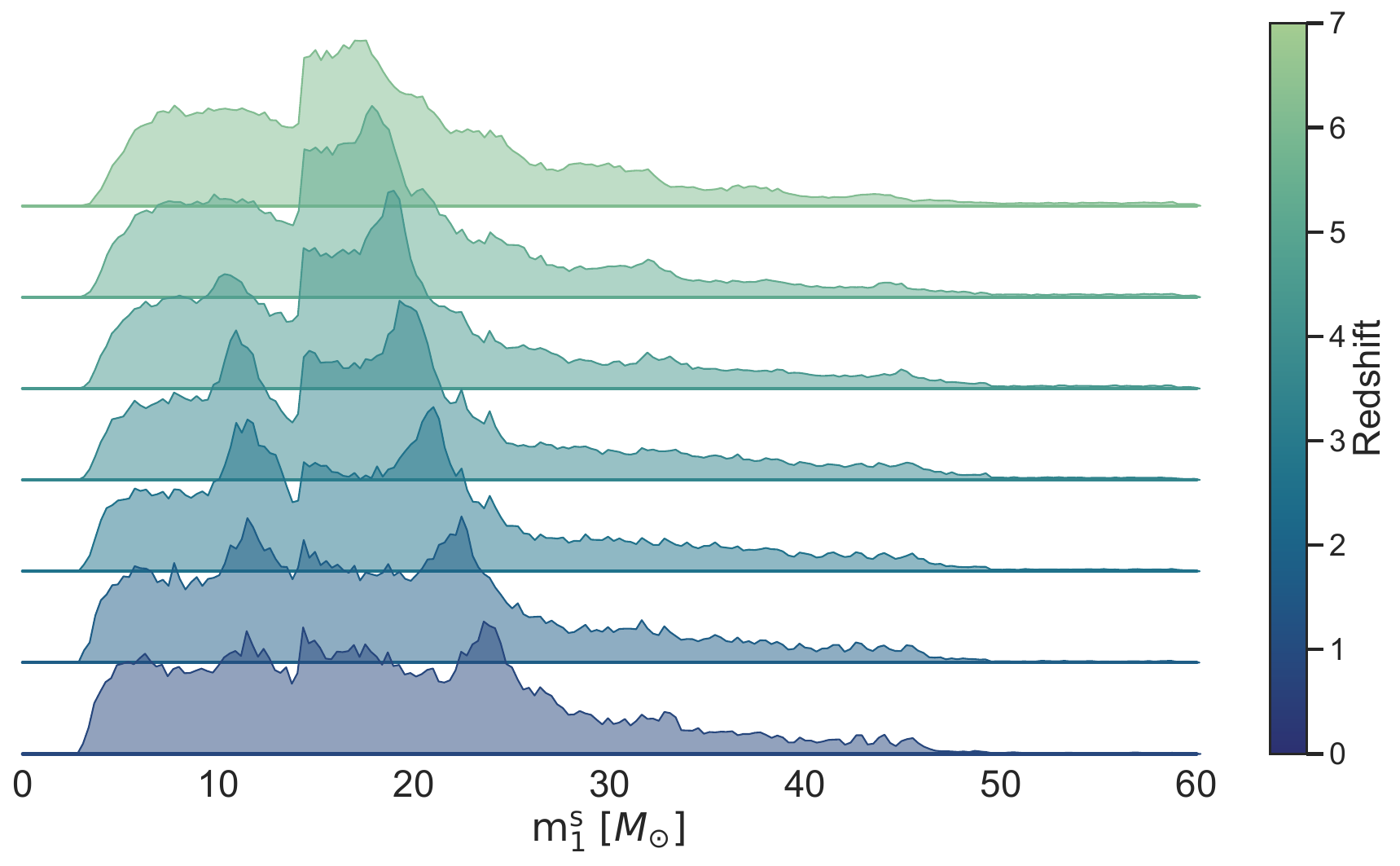}
    \caption{Kernel density estimation of the primary masses $\rm m_1$ in source frame from the A03 BBH catalog. Each plot corresponds to a slice in redshift, going from $z=0$ up to $z=7$. This figure highlights the overall mass spectrum evolution with respect to the redshift, especially the appearance of peaks and their shift.}
    \label{fig:mass spec evolution z A03}
\end{figure}
\begin{figure}
    \centering
    \includegraphics[scale=0.29]{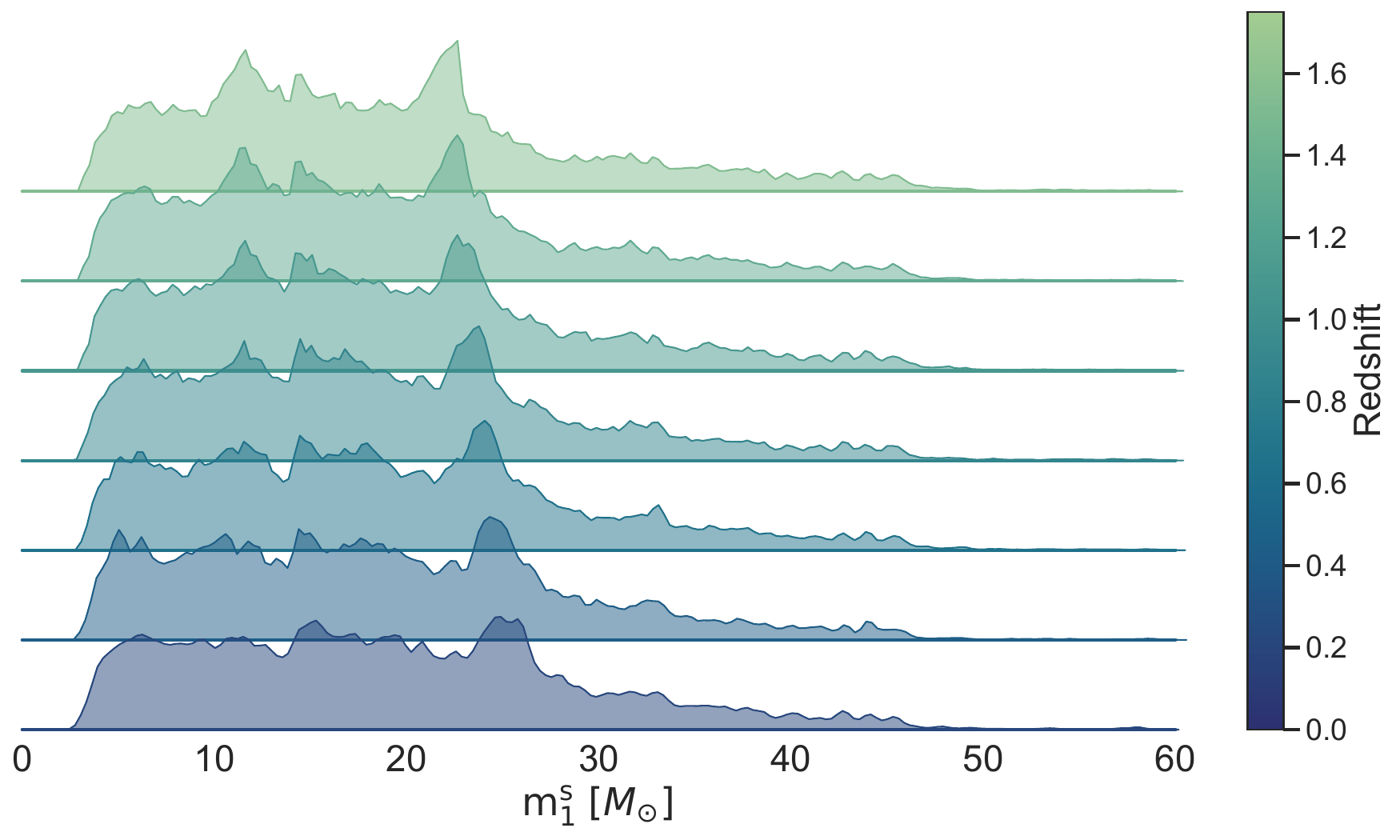}
    \caption{Kernel density estimation of the primary masses $\rm m_1$ in source frame from the A03 BBH catalog. Each plot corresponds to a slice in redshift, going from $z=0$ up to $z=1.75$. This figure highlights the overall mass spectrum evolution with respect to the redshift, especially the appearance of peaks and their shift.}
    \label{fig:mass_spec_evolution_z_low A03}
\end{figure}

Fig.~\ref{fig:PPC linear zevol} summarizes the PPC for this test case. The right panel shows the detected mass distributions for three of the seven simulations. In the detector frame, the 3 simulations display similar distributions for the masses and luminosity distance of detected GW events. The distributions are the ones that are fit by the non-evolving PLP models. However, in the source frame, the three simulations have a significantly different mass spectrum (due to redshift evolution). The PLP model is not able to catch this evolution and produces a wrong reconstruction of the mass spectrum. In particular, the non-evolving PLP model always reconstructs a peak located at 30 $\Msol$. This is the consequence of the fact that about 80\% of the GW detections are located at low redshift ($z<0.3$), where even for redshift evolving models, the peak of the Gaussian component is located around $\mu_{\rm g}^{0}=30 \Msol$.

The introduction of a $H_0$ bias is due to the misplacing of the Gaussian component for events at higher redshifts. When $\mu_{\rm g}^{1}>\mu_{\rm g}^{0}$ and the PLP is not able to fit it, we obtain a $H_0$ biased to higher values. The motivation for this bias is that, for high $H_0$ values, GW events are placed at higher redshifts and hence at lower source masses. By placing the events at lower source masses, the PLP model is able to accommodate them in the peak that fits around $30 \Msol$ in the source frame. When $\mu_{\rm g}^{1}<\mu_{\rm g}^{0}$, we observe a bias to lower values of $H_0$. Similarly to the previous case, this is due to the fact that GW events are placed at lower redshifts, and therefore they will have a higher source mass that can be place in the peak at $30 \Msol$. This simple test case shows that the evolution in redshift of a feature in the mass spectrum could potentially bias the GW-based $H_0$ estimation.
\begin{figure}
    \centering
    \includegraphics[scale=0.26]{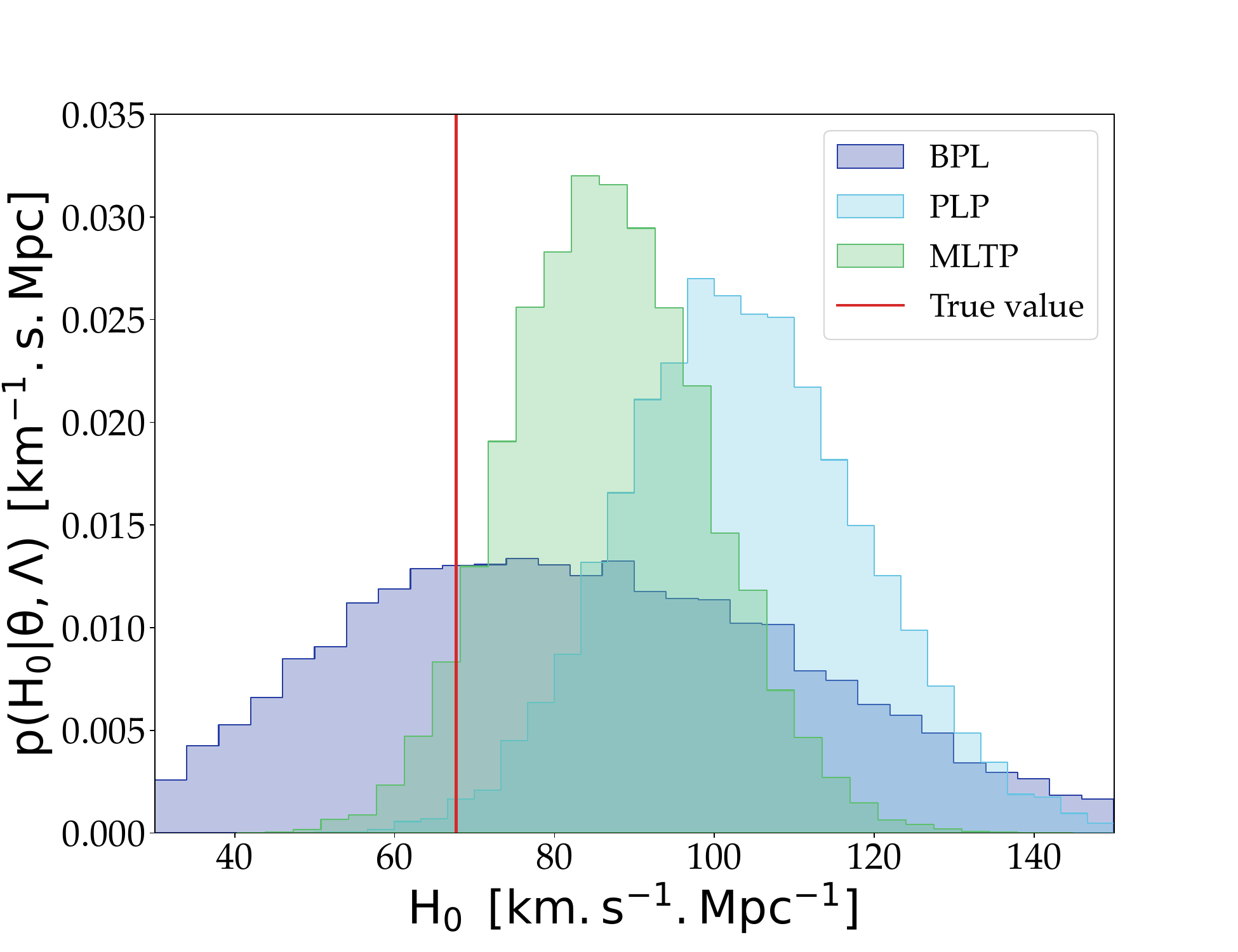}
    \caption{Marginal posterior distributions of the Hubble constant $H_0$ obtained with 2000 GW events, using the PLP in light blue, the BPL in blue and the MLTP mass model in green. The vertical red line represent the true value of the Hubble constant intrinsic to the A03 catalog and equal to $H_0=67.7$ $\hu$.}
    \label{fig:H0 bias A03}
\end{figure}

\subsection{Summary of the tests}

To summarize, the $H_0$ bias is closely linked to the ability of the mass model to catch and reconstruct the features of the CBC mass spectrum. A bias on the $H_0$ value can be introduced from several factors.

We have shown that in the case where the true underlying population of CBCs presents sharp features, and the reconstruction model is not able to include them, a significant bias on $H_0$ can be obtained. We have also shown that if any of the mass features of the spectrum evolve in redshift, a $H_0$ bias can be introduced even if the evolution is of the order of a few solar masses.

\section{\label{Sec4}Application to a complex BBH population}

In this section, we study how the BPL, PLP and MLTP models respond to a population of BBHs generated from synthetic astrophysical simulations. We use a catalogue, referred to as \textit{A03}, of BBH mergers formed by various formation channels from \cite{Mapelli:2021syv, Mapelli:2021gyv}. We then present the results of the hierarchical Bayesian analysis obtained with the BPL, PLP and MLTP mass models in Sec.~\ref{Sec4b}. Finally, in Sec.~\ref{Sec4c}, we investigate the possible sources of bias for the Hubble constant.

\subsection{\label{Sec4a} General description of the A03 catalog}

For this study we employ the "A03" catalog from \cite{Mapelli:2021gyv}. This model consists of four different channels: isolated binary evolution, dynamical assembly of BBHs in young, globular, and nuclear star clusters. They are mixed based on their corresponding redshift-dependent rates as described in \cite{Mapelli:2021gyv}. The dynamical formation channels assume an initial black hole mass function obtained with the MOBSE population synthesis code \cite{Giacobbo:2017qhh}. In model A03, we assume the delayed model by \cite{Fryer:2011cx} for core-collapse supernovae, the pair instability supernova formalism by \cite{Mapelli:2019ipt}, and we take into account the dependence of the BH mass function on the metallicity of the progenitor stars, as discussed in \cite{Giacobbo:2017qhh}. The BHs then pair up dynamically with other BHs and harden because of dynamical interactions in their host star clusters \cite{Mapelli:2021syv}; we also allow for hierarchical mergers of BBHs, as described in  \cite{Mapelli:2021gyv}.
The isolated binary evolution channel  consists of systems evolved with the MOBSE binary population synthesis code accounting for binary star evolution \cite{Giacobbo:2017qhh}. For these runs, we include a treatment for mass transfer, common envelope (with efficiency parameter $\alpha=1$), tidal evolution, natal kicks, and gravitational-wave decay as described in \cite{Giacobbo:2017qhh}. A further discussion of the astrophysical processes included in model A03 and the corresponding uncertainties are beyond the scope of this study. Here, we explore the possibility that non-trivial structures of the mass distribution for BBHs can introduce a bias on the estimation of $H_0$.

\begin{figure*}[htb!]
    \centering
    \includegraphics[scale=0.28]{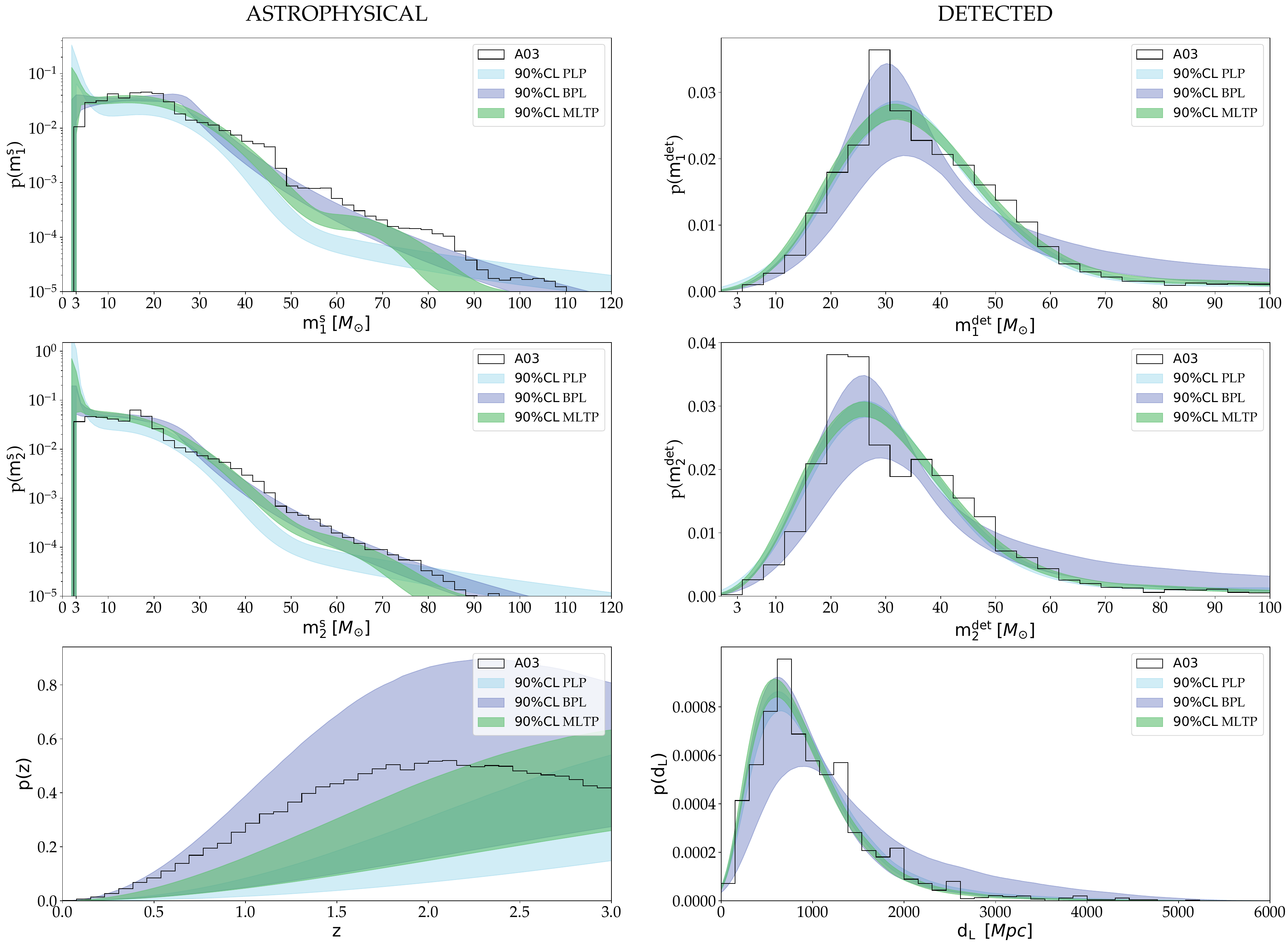}
    \caption{Posterior predictive check for the spectral sirens analysis with 2000 GW observations, generated from the A03 BBH catalog. The plain black histograms are the injected population, in source frame on the left panels and in detector frame on the right panels. The three coloured contours depict the 90\%CL for the three inferences (BPL, PLP and MLTP).}
    \label{fig:PPC A03 vanilla}
\end{figure*}

Fig.~\ref{fig:mass spec evolution z A03} shows the distribution of the primary mass $\rm m_1^s$ for different redshift bins. The BBH mass spectrum presents several features. Moving from lower to higher masses,  a feature at $10 \Msol$, a ``valley'' between $13-15 \Msol$ and a second large peak at $16 \Msol$. In the high mass region of the spectrum, a local over-density of BBHs is visible around $25 M_{\odot}$, plus several less significant peaks up to $90 \Msol$. These structures are also visible on the $\rm m_2^{s}$ distribution. Another property of the A03 catalog is the evolution in redshift of the primary and secondary source frame masses. From Fig.~\ref{fig:mass spec evolution z A03}, we notice that the structures in the mass spectrum are evolving as the redshift increases. The feature at $20 \Msol$ is nonexistent for mergers below $z=2$, and emerges while shifting to the lower masses in the higher redshift bins. 

Fig.~\ref{fig:mass_spec_evolution_z_low A03} shows the BBH mass spectrum in a redshift range of $0<z<1.75$. This range corresponds to the typical range in which our simulation is able to detect BBH mergers. The evolution of the BBH mass spectrum at low redshifts is fainter than the one at high redshift, nevertheless peaks around $11 \Msol$ appear while shifting to higher masses and the structure around $25\Msol$ is drifting to lower masses. Since real GW events at high redshift ($z>1$) are hardly detectable in the O4 like network, this effect is thought to be subdominant for population and cosmological inferences \cite{KAGRA:2021duu}. If these structures are evolving with respect to the redshift, as discussed in \cite{Belczynski:2001uc, Spera:2017fyx, Mapelli:2017hqk, Vitale:2018yhm, Renzo:2020rzx}, the mass models could mismatch the features and produce a biased value of $H_0$ as shown in Sec.~\ref{Sec2c}. 
\begin{figure*}[htb!]
    \centering
    \includegraphics[scale=0.28]{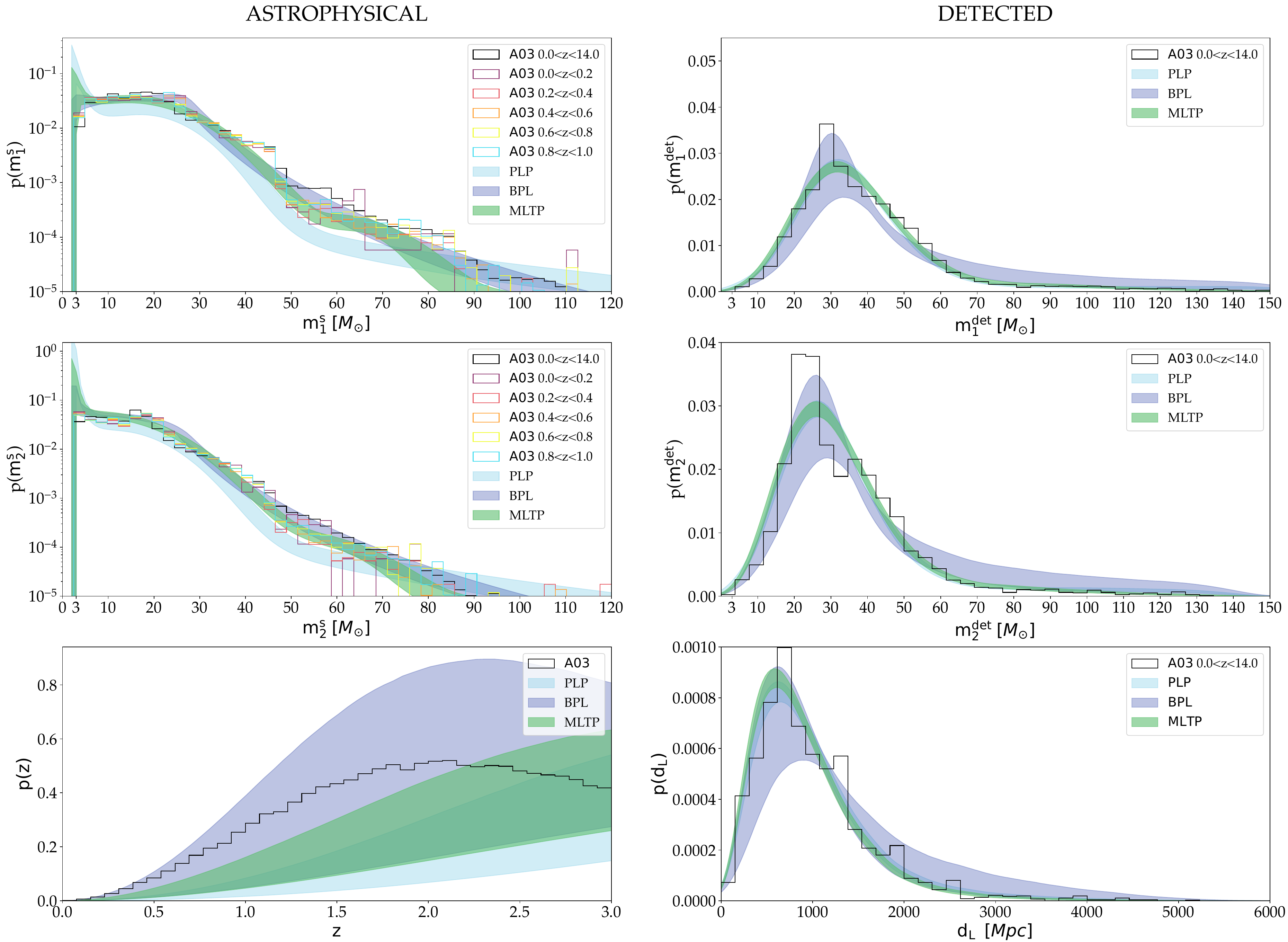}
    \caption{Posterior predictive check of the spectral sirens analysis of 2000 GW observations generated from the A03 BBH catalog. The plain black histograms are the injected population, in source frame on the left panels and in detector frame on the right panels. Each coloured histogram corresponds to a bin in redshift from $z=0$ to $z=2$. The three coloured contours depict the 90\%CL computed with the BPL, PLP and MLTP mass models.}
    \label{fig:PPC A03 vanilla zbinned}
\end{figure*}

\subsection{\label{Sec4b}Vanilla analysis of the A03 catalog}

Using the A03 BBH catalog, and with the framework described in Sec.~\ref{Sec2a}, we generate GW catalogs made of $2000$ detected GW events. For each catalog, we perform a full hierarchical Bayesian inference reconstructing posteriors on $H_0$ as well as the other population parameters of the redshift rate and the three redshift-independent BBH mass models. The prior ranges used for the Bayesian inference are reported in the Tab.~\ref{tab:prior range BPL}, Tab.~\ref{tab:prior range PLP} and Tab.~\ref{tab:prior range MLTP}. 
\begin{figure}
    \centering
    \includegraphics[scale=0.26]{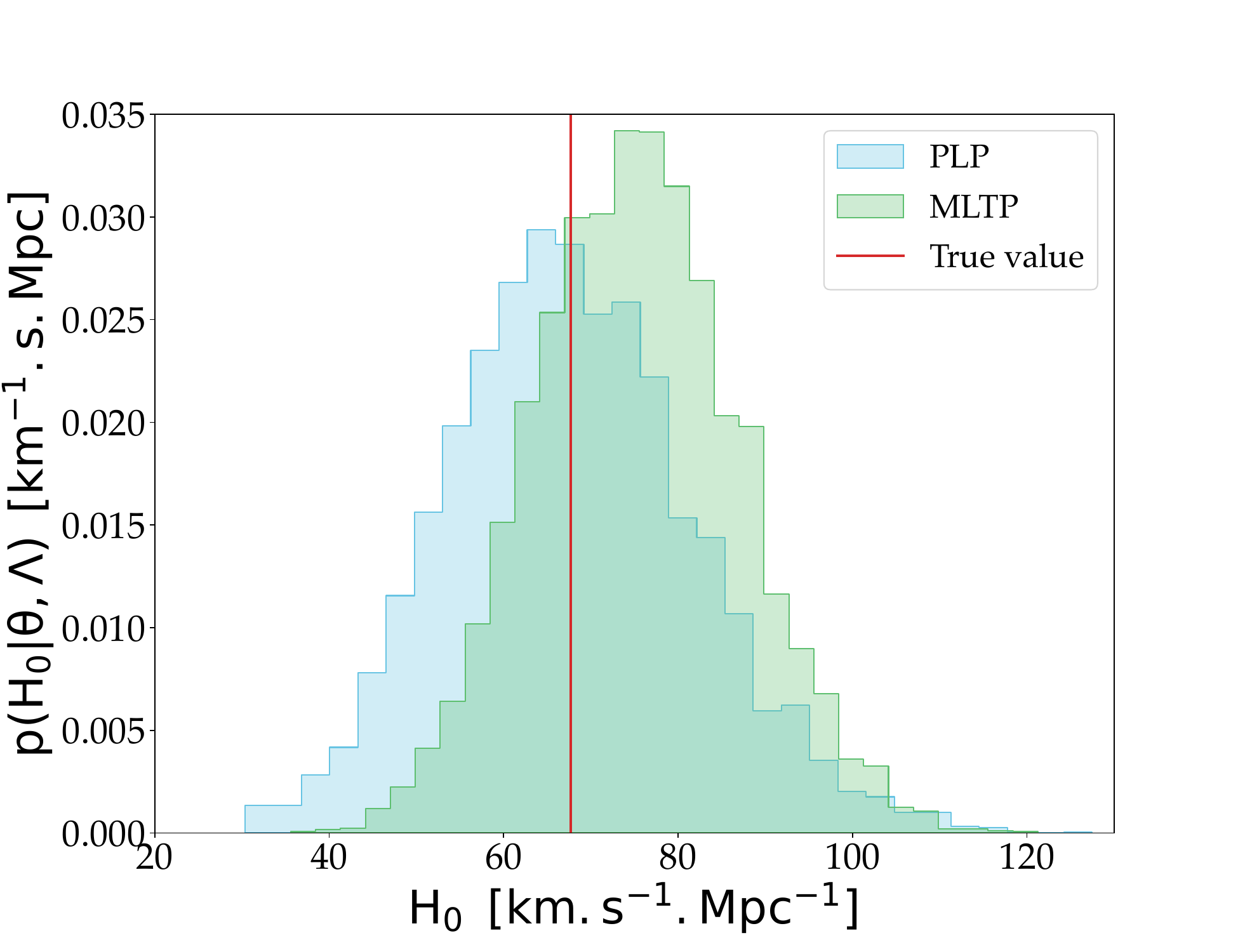}
    \caption{Marginal posterior distributions of the Hubble constant $H_0$ obtained with 2000 GW events, using the PLP and the MLTP mass models. The vertical purple line represent the true value of the Hubble constant intrinsic to the A03 catalog and equal to $H_0=67.7$ $\hu$. This distribution is obtained using the A03 catalog, where the redshift evolution of the mass spectrum has been removed.}
    \label{fig:H0 no evol A03}
\end{figure}

Fig.~\ref{fig:H0 bias A03} shows the marginal posterior distributions of the Hubble constant estimated. When using the PLP and MLTP mass models, we find an $H_0$ bias towards higher values. In particular, we estimate for the PLP $H_0 = 103^{+11}_{-10} \hu$ and $H_0 = 87^{+9}_{-8} \hu$ at 68.3 \% C.I. The true value of $H_0$ is excluded from the 99.7\% C.I. for the PLP and the 95\% C.I.  for the MLTP. From Fig.~\ref{fig:H0 bias A03}, it is possible to note that the BPL mass model obtains an uninformative $H_0$ posterior. We obtain a value of $H_0 = 82^{+20}_{-19} \hu$ which includes the injected $H_0$ within the $68\%$ C.I. As we will argue later, the BPL recovers a less informative posterior on $H_0$ as it does not contain any strong mass feature. We note that, even if the BPL does not display a $H_0$ bias for 2000 events, it might still display it as more and more GW detections are collected. The bias is not introduced by any of the other population parameters ``railing'' over the prior ranges. In fact, we verified that all the parameters are well-constrained in their prior ranges. 

\begin{figure*}[htb!]
    \centering
    \includegraphics[scale=0.28]{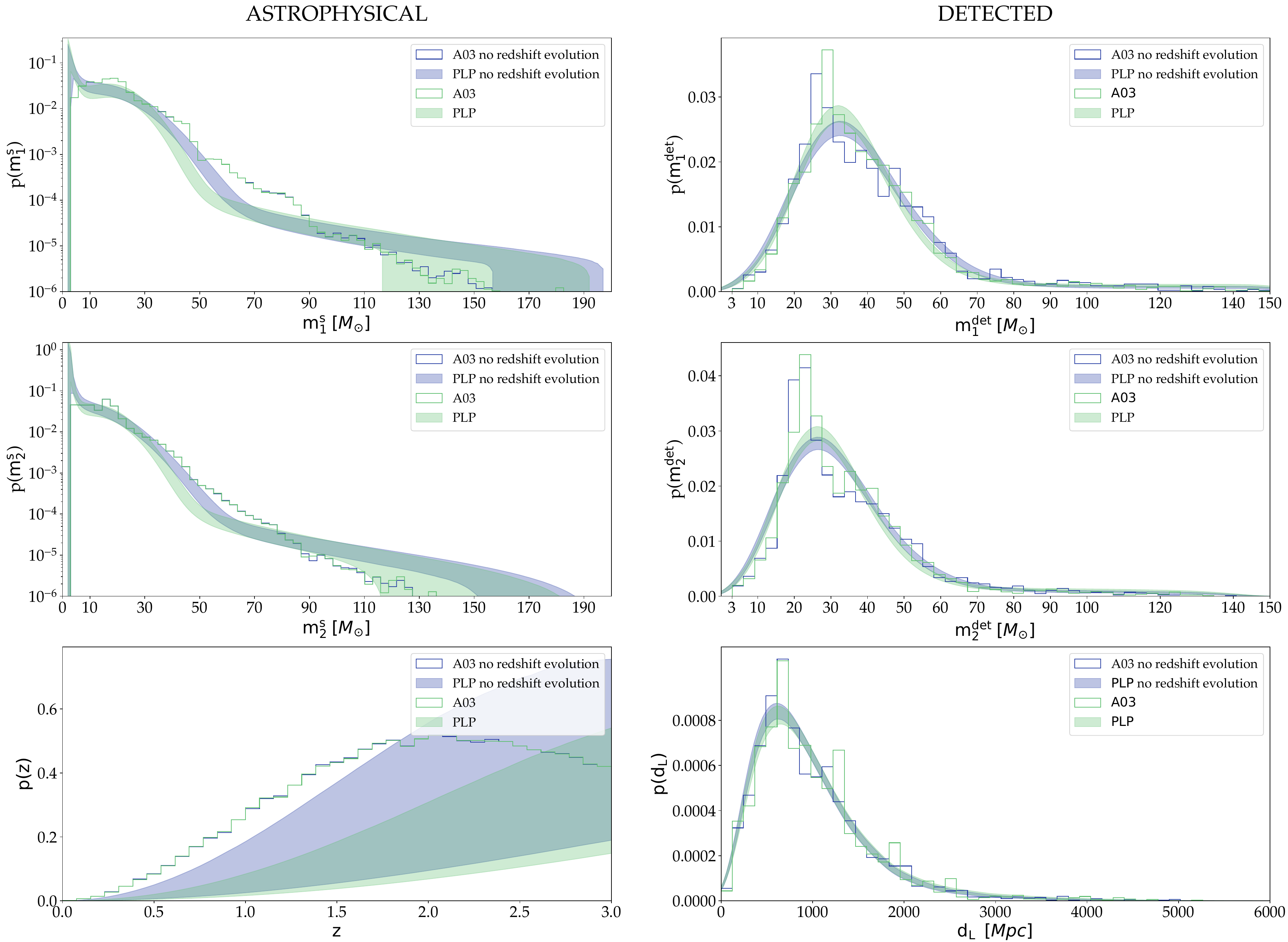}
    \caption{Comparison of posterior predictive checks for the spectral sirens analysis of the A03, with and without redshift evolution of the mass spectrum, using 2000 GW events. The blue contour and histograms are the vanilla population, and are the same results shown in Fig.~\ref{fig:PPC A03 vanilla}. The purple histograms are the new A03 population with no redshift evolution of the mass spectrum and the purple contours are the 90\%CL inferred spectrum with the population, denoted "A03 no redshift evolution".}
    \label{fig:PPC A03 nozevol}
\end{figure*}

The PPC plot in Fig.~\ref{fig:PPC A03 vanilla} gives insights into the origin of the $H_0$ bias. While the reconstructed distributions of detector frame masses and luminosity distance do not show any particular deviation between the injected population and the recovered one, the reconstructed source population display larger differences. In the source frame, the CBC merger rate is correctly reconstructed by the BPL, whereas the PLP and MLTP significantly underestimate the presence of BBHs between $40\Msol$ and $80 \Msol$. Therefore, from observed data, the PLP and MLTP, predict a significantly larger number of BBHs at lower source masses. This is achieved by pushing the value of the Hubble constant at higher values so that events are found at higher redshifts and have on average a lower value of source masses. Another indicator in support of the finding that the PLP and MLTP models place events at higher redshifts is the reconstruction of the CBC merger rate in redshift. In fact, from the bottom left plot of Fig.~\ref{fig:PPC A03 vanilla}, we can see that the PLP and MLTP reconstruct a biased merger rate in redshift that prefers higher values of redshifts for the BBH mergers. Instead, the BPL inference manages to fit the overall trend of the A03 source frame population. The BPL reconstructs better the source frame masses between $40\Msol$ and $80 \Msol$. 

Even though the PLP and MLTP have enough degrees of freedom to approximate a smooth distribution of masses (as the ones present in the A03 catalog), they still fail to reconstruct the mass spectrum. The motivation for which the PLP and MLTP are not able to correctly reconstruct the mass spectrum of the A03 catalog and $H_0$ is not trivial. A first hypothesis for this result was that the PLP and MLTP were catching local peaks in the mass spectrum (at a fixed redshift) that were evolving in redshift. In fact, from the posterior distributions, we observed that the PLP and MLTP were finding local peaks at low masses ($\approx 17 M_{\odot}$) with a large standard deviation and a higher peak at $\approx 30 M_{\odot}$. To test this hypothesis, we performed a PPC in Fig.~\ref{fig:PPC A03 vanilla zbinned} dividing the true population in redshift bins. Unfortunately, the PPC check does not clearly show a discrepancy between the population reconstruction and the any of the mass spectra for the redshift bins. As this test is inconclusive on the origin of the $H_0$ bias, we performed additional studies that we discuss below.

\subsection{\label{Sec4c} Investigating the sources of the $H_0$ bias: blinding the mass-redshift relation}

In order to test further the possibility that the redshift evolution of the mass spectrum is introducing a $H_0$ bias, we performed a simulation where we blind the A03's mass spectra to the redshift evolution. To remove the redshift dependency of the mass spectrum, we randomly shuffle the pairs of BBH merger redshifts and masses. This procedure artificially removes the redshift dependence of the mass spectrum while conserving its non-trivial shape. For the Bayesian inference, we only use the PLP and MLTP mass models since they were displaying a non-negligible bias of $H_0$. 

The result of the joint inference with the PLP mass model is presented in Fig.~\ref{fig:H0 no evol A03}. For the PLP, we obtain $H_0 = 67.0^{+9.7}_{-8.9} \hu$ and for the MLTP we obtain $H_0 = 75.13^{+7.8}_{-7.9} \hu$, the true value of $H_0$ is contained within the 68.3\% C.I. of the posterior. Fig.~\ref{fig:PPC A03 nozevol} shows the PPC plot for this test. The figure overlaps the reconstruction for the original \textit{A03} catalog and the one after the redshift-blinding procedure. From the detector frame point of view, we see that the two reconstructed distributions do not significantly differ. However, from the source frame point of view, the PLP model, for the redshift-blinded \textit{A03} catalog, is able to reconstruct better the BBH masses present in the range $30 \Msol -60 \Msol$.  Although the reconstruction of the mass spectrum above $60 \Msol$ is still poor, the PLP for the non-redshift evolving A03 is able to include the true value of $H_0$. As a consequence, also the reconstruction of the BBH merger rate as a function of redshift slightly improves, resulting in an underestimation of about 20\%-30\% of the true CBC merger rate. We have also found similar results for the MLTP model. We verified by running the simulations with 20 different population realizations that the value of $H_0$ obtainable by the PLP and MLTP from the redshift-blinded A03 is always unbiased.

It is important to notice that, even after blinding the redshift evolution of the A03 mass spectrum, the PLP and MLTP are not yet able to completely reconstruct the source frame distributions. As such, we still expect a systematic bias on $H_0$ to be present and hidden by the large statistical uncertainties that can be seen in 
Fig.~\ref{fig:H0 no evol A03}. Identifying the source of these systematic biases is not trivial, but from the test that we performed, we are sure that the redshift evolution of the mass spectrum is expected to be the dominant source of the $H_0$ bias.

\section{\label{Sec5}Conclusions}

In this work we have extensively studied the interplay between systematic biases on $H_0$ and the reconstruction of the BBH mass spectrum for Spectral Siren Cosmology. We performed several simulations using BBH merger catalogs generated with 3 of the current mass models used in literature and we also considered a catalog of synthetically simulated BBHs.

In Sec.~\ref{Sec3}, we showed that simple phenomenological models that do not contain sharp features could introduce a bias on $H_0$ if the true population of BBHs includes a local over-density of sources. Moreover, we showed that the redshift evolution of source frame mass features can induce a $H_0$ bias if not accounted for in the mass models. In particular, the Hubble constant would act as a free parameter and shift to higher value if the true source mass spectrum is underestimated, hence placing more events at higher redshift, and equivalently decreasing the global source mass spectrum. This results is respectively true if the source mass spectrum is overestimated, creating a shift of $H_0$ to lower values.

In Sec.~\ref{Sec4}, using the A03 BBH catalog, we showed that even if the detector frame is well reconstructed by the inference, we might still get a biased value of the Hubble constant. Deviations up to $2\sigma$ and $3\sigma$ from the injected value were observed for the MLTP and PLP inferences, respectively. We observed that while each model yielded to different populations in the source frame, they produce remarkably similar populations in the detector frame. This discrepancy in the source frame mapping appeared to be related to the response of the models to the GW events, as each model attempted to accommodate to the true population in its own unique manner. Consequently, the Hubble constant is used as a compensatory factor, when necessary, to align the inferred populations with the observed data. Finally, we find that the $H_0$ bias obtained with the PLP and MLTP mass model disappears when the redshift evolution of the source mass spectrum is removed from the population.

In light of these findings, we demonstrate that the most commonly used mass models, namely PLP, BPL, and MLTP, combined with the spectral sirens analysis for GW cosmology, are not suitable choices if the BBH population exhibits even a slight redshift evolution of its mass spectrum. In addition, even if the mass spectrum is not evolving with redshift, the mismatched of sharp structures in the BBH mass distribution can also be responsible of a biased estimation of the Hubble constant. We showed that this kind of assumption can be dangerous for the spectral sirens method and more generally for methods that make use of non evolving parametric models of the mass spectrum since the constraining power on the Hubble constant relies on the ability of the phenomenological mass models to fit the position of structures in the mass spectrum. Given that the true distribution of BBHs in our universe remains unknown and may differ significantly from the current state-of-the-art models, future GW cosmology studies based on the spectral sirens method should incorporate source mass models capable of absorbing effects such as the redshift evolution of the mass spectrum and possessing sufficient degrees of freedom to accommodate unforeseen mass features. Failure to do so may introduce biases in future Hubble constant measurements, especially when dealing with a large number of detected GW events.

\begin{acknowledgments}
The authors are grateful for computational resources provided by the LIGO Laboratory and supported by the National Science Foundation Grants PHY-0757058 and PHY-0823459. We thank Viola Sordini for the helpful comments and discussions on the manuscript.  MM acknowledges financial support from the European Research Council for the ERC Consolidator grant DEMOBLACK, under contract no. 770017, and from the German Excellence Strategy via the Heidelberg Cluster of Excellence (EXC 2181 - 390900948) STRUCTURES.

\end{acknowledgments}

\newpage
\clearpage

\onecolumngrid

\appendix

\section{Mass and redshift population models}
\subsection{\label{App:Source mass model}Source mass models}
In this appendix, we present in more details the three phenomenological source mass models used in this paper for the spectral sirens inference. These models are the current mass models used by the LVK scientific collaboration for cosmology results \cite{LIGOScientific:2020ibl, LIGOScientific:2021usb, KAGRA:2021vkt}. We give the mathematical expressions of priors and probability density functions for the : Power Law plus Peak, the Broken Power Law and the Multi Peak model. Note that they are the same models described in \cite{Mastrogiovanni:2023zbw}. All three models are made of combinations of  Truncated Power Law and Truncated Gaussian distributions.

\noindent The usual truncated Power Law distribution is given by : 
\begin{equation}
    \rm \mathcal{P}(x|x_{min},x_{max},\alpha) = 
    \begin{cases}
        \rm \frac{1}{N} x^{\alpha}, & \left(\rm x_{\rm min}\leqslant x \rm \leqslant x_{\rm max}\right), \\
        \rm 0, & \mathrm{otherwise},
    \end{cases}
    \label{eq:Truncated PL}
\end{equation}
where $\rm N$ is the normalization factor defined as :
\begin{equation}
    \rm N = 
    \begin{cases}
        \rm ln (\frac{x_{\rm max}}{x_{\rm min}}), & \left(\rm  if \hspace{0.15cm} \alpha = -1 \right), \\
        \rm \frac{1}{\alpha +1}(x_{\rm max}^{\alpha+1}-x_{\rm min}^{\alpha+1}), & \mathrm{otherwise}.
    \end{cases}
    \label{eq:Truncated PL normalisation}
\end{equation}

\noindent The Truncated Gaussian distribution with  mean $\mu$ and standard deviation $\sigma$:
\begin{equation}
\mathcal{G}_{[a,b]}(x|\mu,\sigma)=
    \begin{cases}
        \frac{1}{N}\frac{1}{\sigma\sqrt{2\pi}} \exp{\left[ -\frac{(x-\mu)^2}{2\sigma^2}\right]}, & a \leqslant x \rm \leqslant b, \\
        \rm 0, & \mathrm{otherwise},
    \end{cases}
\end{equation}
where $\rm N$ is also the normalization factor expressed as :
\begin{equation}
    \rm N = \int^{\rm a}_{\rm b} \frac{1}{\sigma \sqrt{2 \pi}}e^{-\frac{(x-\mu)^2}{2\sigma^2}}dx.
 \label{eq:truncated gaussian normalisation}
\end{equation}

\noindent \textbf{The Broken Power Law (BPL):}
This model introduced in \cite{LIGOScientific:2020kqk}, is made of two truncated Power Law attached at a breaking point b, 
\begin{equation}
    \rm b = m_{min} + (m_{max}-m_{min})f,
    \label{eq:breaking point b}
\end{equation}
where $f$ belongs to $[0,1]$. When $f=0$, the breaking point b is equal to $\rm m_{min}$. The probability density functions for $\rm m_{1}^{s}$ and $\rm m_{2}^{s}$ is then defined as
\begin{equation}
    \rm \pi(m_{1}^{s}|m_{min},m_{max},\alpha) = \frac{1}{N}\left[\mathcal{P}(m_{1}^s|m_{min},b,-\alpha_1) +\frac{\mathcal{P}(b|m_{min},b,-\alpha_1)}{\mathcal{P}(b|b,m_{max},-\alpha_2)} \mathcal{P}(m_{1}^s|m_{max},b,-\alpha_2) \right],
    \label{eq:pdf m1 BPL}
\end{equation}
\begin{equation}
    \rm \pi(m_{2}^{s}|m_{min},m_{1}^{s},\beta) = \mathcal{P}(m_{2}^s|m_{min},\beta,m_{1}^{s}).
    \label{eq:pdf m2 BPL}
\end{equation}
N is the normalization factor defined as : 
\begin{equation}
    \rm N = 1 + \frac{\mathcal{P}(b|m_{min},b,-\alpha_1)}{\mathcal{P}(b|b,m_{max},-\alpha_2)}.
    \label{eq:norm fac BPL}
\end{equation}
Tab.~\ref{tab:prior range BPL} presents the parameters that control the BPL distribution, as well as the prior distributions used to perform the joint inference of the cosmology and the BBH population presented in Sec.~\ref{Sec4}.

\begin{table}[h!]
    \centering
    \begin{tabular}{ c p{10cm} p{2mm} p{4cm} }
    \hline
    \hline
        {\textbf{Parameter}} & \textbf{Description} &  & \textbf{Prior} \\\hline\hline
        $\alpha_1$ & Power Law index number 1 primary mass. &  & $\mathcal{U}$($-4$, $10$)\\
        $\alpha_2$ & Power Law index number 2 primary mass. &  & $\mathcal{U}$($-4$, $10$)\\
        $\beta$ & Power Law index secondary mass. &  & $\mathcal{U}$($-4, 10$)\\
        $\rm m_{min}$ &  Minimum value of the source mass $[\Msol]$. &  & $\mathcal{U}$($1 \Msol$, $10 \Msol$)\\
        $\rm m_{max}$ & Maximum value of the source mass $[\Msol]$. &  & $\mathcal{U}$($100\Msol$, $200\Msol$) \\
        $\delta_{\rm m}$ & Smoothing parameter $[\Msol]$.  &  & $\mathcal{U}$($0 \Msol$, $10 \Msol$) \\
        $\rm b$ & Breaking point $[\Msol]$.  &  & $\mathcal{U}$($0 \Msol$, $1 \Msol$)\\
        \hline 
        \hline 
    \end{tabular}
    \caption{Summary of the prior ranges used with the BPL mass models for the spectral sirens analysis.}
  \label{tab:prior range BPL}
\end{table}

\noindent \textbf{The Power Law plus Peak (PLP):}
This model composed of a Truncated Power Law and a Gaussian component was first introduced by \cite{Talbot:2019okv}. The probability density functions of the primary masses are given by : 
\begin{equation}
    \rm \pi(m_{1}^s|m_{min},m_{max},\alpha) = (1-\lambda)\mathcal{P}(m_{1}^{s}|m_{min},m_{max},-\alpha) +\lambda \mathcal{G}(m_1^s|\mu_g,\sigma_g), (0\leqslant \lambda \rm \leqslant 1),
    \label{eq:m1  PLP pdf}
\end{equation}
\begin{equation}
    \rm \pi(m_{2}^s|m_{min},m_{max},\alpha) = \mathcal{P}(m_{2}^{s}|m_{1}^s,m_{min},\beta).
    \label{eq:m2  PLP pdf}
\end{equation}
Tab.~\ref{tab:prior range PLP} presents the parameters that control the PLP distribution, as well as the prior distributions used to perform the joint inference of the cosmology and the BBH population presented in Sec.~\ref{Sec4}.
\begin{table}[h!]
    \centering
    \begin{tabular}{ c p{10cm} p{2mm} p{3cm} }
    \hline
    \hline
        {\textbf{Parameter}} & \textbf{Description} &  & \textbf{Prior} \\\hline\hline
        $\alpha$ & Spectral index for the PL of the primary mass distribution. &  & $\mathcal{U}$($1$, $10$)\\
        $\beta$ & Spectral index for the PL of the mass ratio distribution. &  & $\mathcal{U}$($-4$, $10$)\\
        $\rm m_{min}$ & Minimum mass of the primary mass distribution $[\Msol]$. &  & $\mathcal{U}$($1 \Msol$, $12 \Msol$)\\
        $\rm m_{max}$ &  Maximum mass of the primary mass distribution $[\Msol]$. &  & $\mathcal{U}$($50 \Msol$, $200 \Msol$)\\
        $\lambda_{\rm g}$ & Fraction of the model in the Gaussian component. &  & $\mathcal{U}$($0$, $1$) \\
        $\mu_{\rm g}$ & Mean of the Gaussian in the primary mass distribution $[\Msol]$.  &  & $\mathcal{U}$($10 \Msol$, $40 \Msol$) \\
        $\sigma_{\rm g}$ & Width of the Gaussian  in the primary mass distribution $[\Msol]$.  &  & $\mathcal{U}$($6 \Msol$, $17 \Msol$)\\
        $\delta_{m}$ & Range of mass tapering at the lower end of the mass distribution $[\Msol]$.  &  & $\mathcal{U}$($0 \Msol$, $12 \Msol$)\\
        \hline 
        \hline 
    \end{tabular}
    \caption{Summary of the prior ranges used with the PLP mass models for the spectral sirens analysis.}
  \label{tab:prior range PLP}
\end{table}

\noindent \textbf{The Multi Peak (MLTP):}
This model is an extension of the PLP model, it consists of a Truncated Power Law and two Gaussian components. This parameterization was first applied in \cite{LIGOScientific:2020kqk} and the primary masses PDF are given by : 
\begin{equation}
    \rm \pi(m_{1}^s|m_{min},m_{max},\alpha) = \left[  (1-\lambda)\mathcal{P}(m_1^s|m_{min},m_{max},-\alpha) + \lambda \lambda_{low}\mathcal{G}(m_1^s|\mu_g^{low},\sigma_{g}^{low}) +\lambda(1-\lambda_{low})\mathcal{G}(m_1^s|\mu_g^{high},\sigma_{g}^{high})\right],
    \label{eq:m1  MLTP pdf}
\end{equation}
\begin{equation}
    \rm \pi(m_{2}^s|m_{min},m_{max},\beta) = \mathcal{P}(m_2^s|m_{1}^s,m_{min},\beta).
    \label{eq:m2  MLTP pdf}
\end{equation}
Tab.~\ref{tab:prior range MLTP} presents the parameters that control the MLTP distribution, as well as the prior distributions used to perform the joint inference of the cosmology and the BBH population presented in Sec.~\ref{Sec4}.
\begin{table}[h!]
    \centering
    \begin{tabular}{ c p{10cm} p{2mm} p{4cm} }
    \hline
    \hline
        {\textbf{Parameter}} & \textbf{Description} &  & \textbf{Prior} \\\hline\hline
        $\alpha$ & Power Law index primary mass. &  & $\mathcal{U}$($1$, $10$)\\
        $\beta$ & Power Law index secondary mass. &  & $\mathcal{U}$($-4$, $10$)\\
        $\rm m_{min}$ &  Minimum value of the source mass $[\Msol]$. &  & $\mathcal{U}$($1 \Msol$, $10 \Msol$)\\
        $\rm m_{max}$ & Maximum value of the source mass $[\Msol]$. &  & $\mathcal{U}$($50 \Msol$, $200 \Msol$) \\
        $\delta_{\rm m}$ & Smoothing parameter $[\Msol]$.  &  & $\mathcal{U}$($0 \Msol$, $10\Msol$) \\
        $\rm \mu_{g}^{low}$ & Mean of the lower gaussian peak $[\Msol]$.  &  & $\mathcal{U}$($11 \Msol$, $30 \Msol$) \\
        $\rm \mu_{g}^{high}$ & Mean of the higher gaussian peak $[\Msol]$.  &  & $\mathcal{U}$($40 \Msol$, $80 \Msol$) \\
        $\rm \sigma_{g}^{low}$ & S.t.d of the higher gaussian peak $[\Msol]$.  &  & $\mathcal{U}$($6 \Msol$, $17 \Msol$) \\
        $\rm \sigma_{g}^{high}$ & S.t.d of the higher gaussian peak $[\Msol]$.  &  & $\mathcal{U}$($6 \Msol$, $17 \Msol$) \\
        $\rm \lambda_{g}$ & Proportion of events in the peaks.  &  & $\mathcal{U}$($0 $, $1 $) \\
        $\rm \lambda_{g}^{low}$ & Proportion of events in the lower peak.  &  & $\mathcal{U}$($0 , 1$) \\
        \hline 
        \hline 
    \end{tabular}
    \caption{Summary of the prior ranges used with the MLTP mass models for the spectral sirens analysis.}
  \label{tab:prior range MLTP}
\end{table}

All of the three mass models described above are normalized functions, since they are all made of normalized PDF.

\subsection{\label{App:Merger rate model}Merger rate evolution model}
In this paper, we use a specific parameterization of the CBC merger rate $\mathcal{R}(z)$, also denoted $p_{pop}(z|\Lambda)$ in Sec.~\ref{Sec2} as a function of the redshift. This model is constructed after the Madau $\&$ Dickinson star formation rate \cite{Madau:2014bja}. Using the same notation as in Eq.~\ref{eq:cbc rate full}, the CBC merger rate is expressed as : 
\begin{equation}
    \rm p_{pop}(z|\Lambda) =  \left[1 + (1+z_p)^{-\gamma -k}\right]\frac{(1+z)^{\gamma}}{1+\left(\frac{1+z}{1+z_p} \right)^{\gamma +k}}.
    \label{eq:M&D merger rate}
\end{equation}
Tab.~\ref{tab:prior range BPL} presents the parameters that control the CBC merger rate distribution, as well as the prior distributions used to perform the joint inference of the cosmology and the BBH population presented in Sec.~\ref{Sec4}.
\begin{table}[h!]
    \centering
    \begin{tabular}{ c p{11cm} p{2mm} p{3cm} }
        \hline
        \hline
        {\textbf{Parameter}} & \textbf{Description} &  & \textbf{Prior} \\\hline\hline
        $\gamma$ & Slope of the power law regime for the rate evolution before the point $z_p$ &  & $\mathcal{U}$($0$, $10$) \\
        $k$ & Slope of the power law regime for the rate evolution after the point $z_{\rm p}$ &  & $\mathcal{U}$($0$, $6$)\\
        $z_p$ & Redshift turning point between the power law regimes with $\gamma$ and $k$ &  & $\mathcal{U}$($0$, $4$) \\
        \hline
        \hline 
    \end{tabular}
    \caption{
    Summary of the prior ranges used with the CBC merger rate for the spectral sirens analysis.}
  \label{tab:prior cbc merger rate}
\end{table}

\section{\label{App:BBH population}Injected BBHs populations}
This appendix presents the values for each of the mass and CBC merger rate parameters, chosen to perform the spectral sirens analysis of Sec.~\ref{Sec3}, for each of the following mass models : BPL, PLP and MLTP.
\subsection{\label{App:mass pop injected}Mass parameters simulated}
\subsubsection{\label{App:BPL injected}Broken power law parameters}
\begin{table}[h!]
    \centering
    \begin{tabular}{ c p{10cm} p{2mm} p{4cm} }
    \hline
    \hline
        {\textbf{Parameter}} & \textbf{Description} &  & \textbf{Injected value} \\\hline\hline
        $\alpha_1$ & Power Law index number 1 primary mass. &  & 1.5\\
        $\alpha_2$ & Power Law index number 2 primary mass. &  & 5.5\\
        $\beta$ & Power Law index secondary mass. &  & 1.4\\
        $\rm m_{min}$ &  Minimum value of the source mass $[\Msol]$. &  & $5 \Msol$\\
        $\rm m_{max}$ & Maximum value of the source mass $[\Msol]$. &  & $100 \Msol$ \\
        $\delta_{\rm m}$ & Smoothing parameter $[\Msol]$.  &  & $5 \Msol$ \\
        $\rm b$ & Breaking point $[\Msol]$.  &  & $0.4 \Msol$\\
        \hline 
        \hline 
    \end{tabular}
    \caption{Summary of the prior ranges used with the BPL mass models for the spectral sirens analysis.}
  \label{tab:BPL injected}
\end{table}

\subsubsection{\label{App:PLP injected}Power Law plus peak parameters}
\begin{table}[htb!]
    \centering
    \begin{tabular}{ c p{10cm} p{2mm} p{3cm} }
    \hline
    \hline
        {\textbf{Parameter}} & \textbf{Description} &  & \textbf{Injected value} \\\hline\hline
        $\alpha$ & Spectral index for the PL of the primary mass distribution. &  & 2\\
        $\beta$ & Spectral index for the PL of the mass ratio distribution. &  & 1\\
        $\rm m_{min}$ & Minimum mass of the primary mass distribution $[\Msol]$. &  & $5 \Msol$\\
        $\rm m_{max}$ &  Maximum mass of the primary mass distribution $[\Msol]$. &  & $100\Msol$\\
        $\lambda_{\rm g}$ & Fraction of the model in the Gaussian component. &  & 0.1 \\
        $\mu_{\rm g}$ & Mean of the Gaussian in the primary mass distribution $[\Msol]$.  &  & $35\Msol$ \\
        $\sigma_{\rm g}$ & Width of the Gaussian  in the primary mass distribution $[\Msol]$.  &  & $5\Msol$\\
        $\delta_{m}$ & Range of mass tapering at the lower end of the mass distribution $[\Msol]$.  &  & $5\Msol$\\
        \hline 
        \hline 
    \end{tabular}
    \caption{Summary of the values injected to construct the PLP mass distribution of the BBH populations analysed in Sec.~\ref{Sec3}.}
  \label{tab:PLP injected}
\end{table}

\subsubsection{\label{App:MLTP injected}Multi Peak parameters}

\begin{table}[htb!]
    \centering
    \begin{tabular}{ c p{10cm} p{2mm} p{4cm} }
    \hline
    \hline
        {\textbf{Parameter}} & \textbf{Description} &  & \textbf{Injected value} \\\hline\hline
        $\alpha$ & Power Law index primary mass. &  & 2\\
        $\beta$ & Power Law index secondary mass. &  & 1\\
        $\rm m_{min}$ &  Minimum value of the source mass $[\Msol]$. &  & $5\Msol$\\
        $\rm m_{max}$ & Maximum value of the source mass $[\Msol]$. &  & $100\Msol$\\
        $\delta_{\rm m}$ & Smoothing parameter $[\Msol]$.  &  & $5\Msol$\\
        $\rm \mu_{g}^{low}$ & Mean of the lower gaussian peak $[\Msol]$.  &  & $16\Msol$ \\
        $\rm \mu_{g}^{high}$ & Mean of the higher gaussian peak $[\Msol]$.  &  & $50 \Msol$ \\
        $\rm \sigma_{g}^{low}$ & S.t.d of the higher gaussian peak $[\Msol]$.  &  & $8\Msol$ \\
        $\rm \sigma_{g}^{high}$ & S.t.d of the higher gaussian peak $[\Msol]$.  &  & $8\Msol$ \\
        $\rm \lambda_{g}^{high}$ & Proportion of events in the peaks.  &  & 0.7\\
        $\rm \lambda_{g}^{low}$ & Proportion of events in the lower peak.  &  & 0.8 \\
        \hline 
        \hline 
    \end{tabular}
    \caption{Summary of the values injected to construct the MLTP mass distribution of the BBH populations analysed in Sec.~\ref{Sec3}.}
  \label{tab:MLTP injected}
\end{table}

\subsection{\label{App:rate pop injected}Rate parameters simulated}

\begin{table}[htb!]
    \centering
    \begin{tabular}{ c p{11cm} p{2mm} p{3cm} }
        \hline
        \hline
        {\textbf{Parameter}} & \textbf{Description} &  & \textbf{Injected value} \\\hline\hline
        $\gamma$ & Slope of the power law regime for the rate evolution before the point $z_p$ &  & 3 \\
        $k$ & Slope of the power law regime for the rate evolution after the point $z_{\rm p}$ &  & 3\\
        $z_p$ & Redshift turning point between the power law regimes with $\gamma$ and $k$ &  & 2 \\
        \hline
        \hline 
    \end{tabular}
    \caption{
    Summary of the values injected to construct the CBC merger rate of the BBH populations analysed in Sec.~\ref{Sec3}.}
  \label{tab:Rate injected}
\end{table}

\newpage
\clearpage
\twocolumngrid

\nocite{*}

\bibliography{apssamp}

\end{document}